\documentclass[lettersize,journal,dvipsnames]{IEEEtran}

\usepackage{url}

\usepackage{booktabs} 
\usepackage{colortbl} 
\usepackage{arydshln} 
\usepackage{lscape}
\usepackage{cite}
\usepackage{amsmath,amssymb,amsfonts} 
\usepackage{graphicx}
\usepackage{textcomp}
\usepackage{xcolor}
\usepackage{mdframed} 
\usepackage{wasysym}  
\usepackage{graphbox} 
\usepackage{hanging} 

\usepackage{tikz}
\usetikzlibrary{shapes,arrows.meta,positioning,calc}

\graphicspath{{./img/}}

\begin{document}

\title{Closing the Loop in Affect-Driven Game Adaptation: A Systematic Review}

\author{Phil~Lopes,~\IEEEmembership{Member,~IEEE,}
        Nuno~Fachada,
        and~Micaela~Fonseca
\thanks{This research was partially funded by: Fundação para a Ciência e a Tecnologia (FCT, \protect\url{https://ror.org/00snfqn58}) under grants
UID/05380/2025 (\protect\url{https://doi.org/10.54499/UID/05380/2025}), 
UID/PRR/05380/2025 (\protect\url{https://doi.org/10.54499/UID/PRR/05380/2025}), 
UID/PRR2/05380/2025 (\protect\url{https://doi.org/10.54499/UID/PRR2/05380/2025}), 
UID/06486/2025 (\protect\url{https://doi.org/10.54499/UID/06486/2025}), 
UID/PRR/06486/2025 (\protect\url{https://doi.org/10.54499/UID/PRR/06486/2025}), 
UID/PRR2/06486/2025 (\protect\url{https://doi.org/10.54499/UID/PRR2/06486/2025}), 
and CEECINST/00002/2021/CP2788/CT0001
(\protect\url{https://doi.org/10.54499/CEECINST/00002/2021/CP2788/CT0001}); 
and Instituto Lusófono de Investigação e Desenvolvimento (ILIND, \protect\url{https://ror.org/02qy8ba98}) under Project COFAC/ILIND/COPELABS/1/2024. 
(Corresponding author: Phil Lopes.)}%
\thanks{Phil Lopes and Micaela Fonseca are with HEI-Lab, Lusófona University, Campo Grande, 376, 1749-024 Lisboa, Portugal (email: phil.lopes@ulusofona.pt; micaela.fonseca@ulusofona.pt).}%
\thanks{Nuno Fachada is with Lusófona University, Campo Grande, 376, 1749-024 Lisboa, Portugal, and with the INESC INOV-Lab, Rua Alves Redol, 9, 1000-029 Lisboa, Portugal (email: nuno.fachada@ulusofona.pt).}%
}

\markboth{Preprint, April~2026}%
{Lopes \MakeLowercase{\textit{et al.}}: Closing the Loop in Affective Game Adaptation: A Systematic Review}

\maketitle

\begin{abstract}
Recognizing player state is only one component of affective game adaptation; inferred experience must also be translated into adaptive interventions that modify gameplay or game content. Although player experience modeling and content adaptation are established research areas, fewer studies examine how sensing, modeling, and adaptation are integrated into complete, empirically evaluated gameplay systems. This PRISMA-guided systematic review analyzes 23 empirical studies published from January 1, 2015, to December 31, 2025, that implement a complete experience-driven loop defined here as the combination of player data acquisition, player experience modeling, and adaptive game content. Complete-loop systems were relatively uncommon in the retrieved corpus, and the selected systems were predominantly oriented toward dynamic difficulty adjustment, engagement, rehabilitation, or performance-related goals. Game telemetry was the dominant input modality, while non-invasive sources with affective relevance, such as facial expression analysis and peripheral interaction data, were less common. Knowledge-based methods, including rule-based systems and heuristics, dominated both modeling and adaptation because of their interpretability and low deployment requirements, whereas machine learning approaches were less frequent and remained constrained by data availability, transparency, and runtime integration challenges. Most importantly, affective information was often used to support challenge calibration or related adaptation objectives, while stress, anxiety, horror, and related affective states were rarely addressed as explicit adaptation targets. These findings identify a gap within this review scope: affective information may enter an adaptive loop without making affective state the objective of adaptation.
\end{abstract}


\IEEEpeerreviewmaketitle

\section{Introduction}
\label{sec:intro}
\IEEEPARstart{A}{ffective} computing faces a central challenge: not only to infer human affective states, but to translate those inferences into interventions that shape subsequent experience. Digital games make this problem especially visible because player experience unfolds through continuous interaction, and affective, behavioral, and physiological signals can be used to modify gameplay or game content in real time. A system may use stress, arousal, frustration, or physiological measures while adapting primarily for difficulty, performance, engagement, or task calibration. However, the presence of affective input does not by itself establish affective adaptation.

Game adaptation research provides a concrete setting in which this distinction can be examined. Since the work of Hunicke~\cite{hunicke2005case}, the idea of adapting games to individual players has grown into an active area of research~\cite{robinson2020letsget,yannakakis2023affective,mortazavi2024dynamic}. One of the most established examples is Dynamic Difficulty Adjustment (DDA), in which game difficulty is modified in response to player performance or state. Affective computing raises a related but distinct question: whether emotional and physiological cues, such as frustration, anxiety, arousal, or stress, are used merely to support gameplay balancing or to drive adaptation aimed at shaping affective experience itself. Yet it remains unclear how often such affective signals are integrated into complete, empirically evaluated systems where player sensing, experience modeling, and content adaptation operate as a complete experience-driven loop.

As pointed out by Yannakakis and Togelius~\cite{yannakakis2011experience}, these approaches can be framed as a user experience problem, where the goal is to optimize player interaction by dynamically modifying game content. Although originally focused on entertainment, these systems have expanded into serious applications such as rehabilitation~\cite{rodrigues2024participation,fonseca2025games}, mental health~\cite{fleming2017serious}, or cognitive training~\cite{pallavicini2018video}. For example, DDA principles can be applied to exergames~\cite{darzi2021user}, ensuring exercises remain engaging and effective. Nonetheless, content adaptation extends beyond difficulty control, having been used to optimize various human performance indicators, such as educational learning performance~\cite{chrysafiadi2023fuzzy}, impact of treatment~\cite{proencca2018serious}, or collaborative potential~\cite{chanel2017multiple}. Recent work has even explored emotion-based environmental orchestration, e.g., enemy placement, level layout, or lighting~\cite{liapis2018orchestrating}. These examples show that adaptation may pursue many objectives, but they also make it necessary to distinguish content adaptation in general from adaptation explicitly targeting affective experience.

The key question for this review is therefore not whether player experience can be modeled in isolation, but whether such models are integrated into adaptive gameplay systems. Many approaches focus on experience recognition without specifying how the inferred state drives content adaptation. The methodological variety of existing systems also makes it necessary to compare how complete-loop systems implement sensing, modeling, and adaptation, and whether the adaptation objective is affective experience or another outcome, such as \textit{gameplay balancing}.

\subsection{Motivation and Scope}
This review was initially motivated by experience-driven adaptation in contexts related to stress, anxiety, and horror, where adaptive content can shape immersion and psychological response~\cite{lopes2017modelling}. Preliminary reading indicated that related work often addresses DDA as an adaptation objective~\cite{paraschos2023game}, while reviews of affect-adaptive games usually discuss emotion-aware methods more broadly rather than focusing on empirically evaluated systems that implement the complete experience-driven loop~\cite{croissant2023theories}. The present review therefore treats horror, stress, and anxiety not only as retrieval terms, but as an analytical focus: it asks whether these states appear as explicit adaptation objectives or only as cues supporting other objectives.

As the results show, complete-loop systems that explicitly target stress, anxiety, horror, or closely related affective states were rare within the selected corpus. This absence is not treated as a failure of this review's scope, but rather as a finding within that scope.

\subsection{Contribution and Research Focus}

This paper systematically reviews empirically evaluated complete-loop systems, with particular attention to how affective information is operationalized within the experience-driven loop. The contribution of this review is a scoped synthesis that distinguishes affect-informed adaptation, where affective information supports another objective such as DDA, engagement, rehabilitation, or performance calibration, from affective adaptation, where affective state is itself the explicit adaptation objective. This distinction is used to interpret how often stress, anxiety, horror, and related states function as adaptation objectives within the selected corpus. To guide this analysis, the following research questions are posed:

\begin{description}
    \item[\textbf{RQ1}] To what extent does the retrieved literature implement and
    empirically evaluate the complete experience-driven loop?

    \item[\textbf{RQ2}] How are the selected complete-loop systems characterized in terms of adaptation objectives, input data types, modeling techniques, adaptation methods, and application domains?

    \item[\textbf{RQ3}] Within the selected complete-loop systems, to what degree have stress, anxiety, and closely related affective states been explicitly addressed as adaptation objectives?
\end{description}

In this context, ``addressed as adaptation objectives'' refers to cases where the \textit{adaptive system explicitly aims to influence, regulate, induce, or sustain affective experience}, rather than merely using affective information to support another objective.

The rest of the paper is organized as follows. Section~\ref{sec:concepts} introduces the conceptual framework underlying player experience modeling and adaptation. Section~\ref{sec:methods} details the systematic review methodology used to gather and analyze relevant literature. Section~\ref{sec:results} presents the findings, highlighting key trends in adaptation methods. Section~\ref{sec:discussion} provides a critical discussion of these findings. Section~\ref{sec:conclusion} summarizes the key takeaways from this study. Finally, the appendices provide a glossary of key terms and detailed summaries of the included studies.

\section{Conceptual Framework}
\label{sec:concepts}

Because this review spans game research, affective computing, procedural content generation, and applied game domains, key terms are not always used consistently across the literature. This section defines the terms used for eligibility and analysis: the complete experience-driven loop, player data, player experience model, content adaptation, adaptation objective, affect-informed adaptation, and affective adaptation.

\subsection{Defining the Experience-Driven Loop}
\label{sec:concepts:defloop}

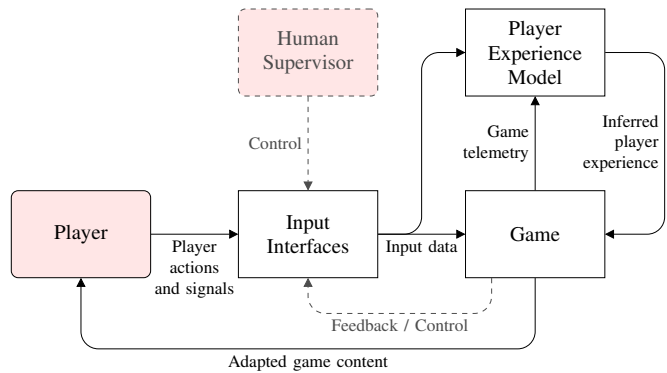
\begin{figure}[t]
    \centering
    \resizebox{\columnwidth}{!}{
    \begin{tikzpicture}[node distance=5.4em and 5.0em]
    \colorlet{optcolor}{black!70}

    \tikzset{
        base/.style={draw, minimum width=8em, minimum height=5em, align=center, font=\large},
        optional/.style={dashed, text=optcolor, draw=optcolor},
        person/.style={base, rounded corners, fill=red!10},
        player/.style={person},
        supervisor/.style={person, optional},
        mech/.style={base},
        arr/.style={->,>={Triangle[length=5pt,width=5pt]},rounded corners=10pt},
      }

    \node[player]     (player)                       {Player};
    \node[mech]       (input)      [right=of player] {Input\\Interfaces};
    \node[supervisor] (supervisor) [above=of input]  {Human\\Supervisor};
    \node[mech]       (game)       [right=of input]  {Game};
    \node[mech]       (pem)        [above=of game]   {Player\\Experience\\Model};

    \draw[arr] (player.east)
    -- node[midway, below, text width=5em, align=center] {Player actions and signals}
    (input.west);

    \draw[arr] (input.east)
    -- node[midway, below] {Input data}
    (game.west);

    \draw[arr] (game.north)
    -- node[midway, left, text width=5em, align=right] {Game telemetry}
    (pem.south);

    \draw[arr] (pem.east)
    -- ++(1.2, 0)
    -- ($(game.east)+(1.2,0)$)
    node[midway, left, text width=5em, align=right] {Inferred player\\experience}
    -- (game.east);

    \draw[arr] (input.east)
    -- ($(input)!0.5!(game)$)
    -- ($(supervisor)!0.5!(pem)$)
    -- (pem.west);

    \draw[arr] (game.south)
    -- ++(0, -1.43)
    -- ($(player.south)+(0,-1.43)$)
    node[midway, below] {Adapted game content}
    -- (player.south);

    \draw[arr, optional] ($(game.south)+(-2.5em,0)$)
    -- ($(game.south)+(-2.5em, -0.715)$)
    -- ($(input.south)+(0, -0.715)$)
    node[midway, below] {Feedback / Control}
    -- (input.south);

    \draw[arr, optional] (supervisor.south)
    -- node[midway, left, align=right] {Control}
    (input.north);

    \end{tikzpicture}
    } 

    \caption{Characterization of the experience-driven loop, based on Yannakakis and Togelius~\cite{yannakakis2011experience}. The player produces observable actions and signals, including behavioral, physiological, and other measurable responses which are captured by input interfaces (e.g., controllers, biosensors, or surveys). The resulting input data and game telemetry are used by the player experience model to infer aspects of the player's experience, which inform subsequent adaptations of game content. Dashed lines represent optional or less common components, such as human supervision or bidirectional communication between the game and input interfaces for purposes such as questionnaire delivery or device calibration.}
    \label{fig:affLoop}

\end{figure}

In this review, a complete experience-driven loop comprises three linked components: player data, a player experience model, and content adaptation. Player data provide input to the model; the model infers a player state or experience relevant to adaptation; and the system uses that inference to modify game content. This characterization, illustrated in Fig.~\ref{fig:affLoop}, follows the experience-driven framework proposed by Yannakakis and Togelius~\cite{yannakakis2011experience} and is used both as an eligibility criterion and as the basis for the coding scheme.

\subsubsection{Player Data}
\label{sec:concepts:defloop:pdata}

Within the scope of this work, \textit{player data} refers to any information used as input to a player experience model, regardless of whether it is collected before, during, or after gameplay. This data provides the foundation upon which adaptive systems build their understanding of player behavior, affective states, or preferences.

To provide a consistent framing for the review and analysis that follows, this work categorizes player data into the following types:

\begin{itemize}
    \item \textbf{Game telemetry:} Includes data derived from the game system, comprising both gameplay metrics (e.g., in-game actions, performance indicators) and game context (e.g., triggered events, scene content). These two forms of data are often intertwined and not easily separable in practice.
    \item \textbf{Physiological data:} Covers biosignals such as electrodermal activity (EDA), heart rate (HR), electromyography (EMG), respiration, electroencephalography (EEG), and related measures.
    \item \textbf{Questionnaires:} Includes self-reported information from players, typically collected pre- or post-play, and intended to reflect preferences, emotional states, or subjective evaluations.
    \item \textbf{Bodily expressions:} Refers to observable expressions of affect through modalities such as speech, facial expressions, body posture, or head movement.
    \item \textbf{Peripheral input:} Low-level interaction signals captured from standard input devices such as keyboards, mice, graphics tablets, touchscreens, or controllers, before or apart from game-level interpretation.
    \item \textbf{Third-person annotations:} Labels or assessments made by external observers or experts regarding the player's behavior or experience.
\end{itemize}

These categories partially reflect the typology proposed by Yannakakis and Togelius~\cite{yannakakis2018artificial}, who distinguish gameplay data, objective data, game context, and static player profile information. For this review, gameplay data and game context are grouped under game telemetry because many selected studies do not report enough detail to separate them reliably.

The broad category of objective data is divided into physiological data, bodily expressions, and peripheral input to reflect differences in sensing modality and relevance to affective or behavioral modeling. Questionnaires and third-person annotations are treated separately because they often provide self-report, observer-based, or contextual labels rather than continuous in-game signals.

\subsubsection{Player Experience Model}
\label{sec:concepts:defloop:pexpmodel}
Within this review, the player experience model refers to any system or process that uses input data to infer a player state or experience relevant to adaptation. This may include affective state, engagement, challenge, performance, preference, or other player-related constructs used to guide content adaptation~\cite{melhart2021towards, hooshyar2018data, kim2017computational}.

Several broad approaches to building such models are commonly found in the literature. Knowledge-based methods, including rule-based systems (e.g., heuristics, if--then adaptation rules), fuzzy logic, and mappings derived from psychological theory are widely used due to their interpretability and low data requirements. These often draw on frameworks such as Csikszentmihalyi’s Flow Model~\cite{csikszentmihalyi1990flow}, which describes the balance between challenge and skill, or Russell’s Circumplex Model of Affect~\cite{russell1980circumplex}, which positions emotions along valence and arousal dimensions.

In contrast, statistical techniques and ML methods offer more data-driven alternatives, capable of identifying patterns and relationships that may be difficult to encode manually. These include both simple models such as regressions or thresholding, and more complex ones such as classifiers or neural networks. While often effective, these approaches may require larger datasets and bring trade-offs in transparency and interpretability. Additionally, some systems rely on proprietary tools, for example commercial emotion recognition hardware, which provide affect estimates using closed-source algorithms.

Although the distinction between these categories is not always strict, it provides a useful basis for coding the modeling techniques used in the selected complete-loop systems. In affective computing, modeling affect remains especially challenging in interactive settings such as games~\cite{wijayarathna2022toward,yannakakis2023affective}.

\subsubsection{Content Adaptation}
\label{sec:concepts:defloop:contadapt}

As the term implies, \textit{content adaptation} refers to the transformation of digital content or game mechanics to optimize a particular player experience or emotional response. While modeling techniques, discussed previously, are concerned with assessing the player's internal state (e.g., challenge, emotion, engagement), content adaptation methods govern how the game system responds to such assessments. This distinction is critical but frequently blurred in the literature, with some works treating modeling and adaptation as a single, undifferentiated process.

Although games are inherently interactive, it is also essential to distinguish between player-driven changes (arising from user input) and system-driven adaptations initiated autonomously by the game. The former reflects immediate, reactive changes based on player actions, while the latter represents a predictive and deliberate system response grounded in player experience modeling. Content adaptation belongs to this latter category.

Content adaptation can be understood through a set of three conceptual levels that describe how it is structured and operationalized in adaptive game systems. At the highest level lies the \textit{adaptation objective}: the intent or goal motivating the adaptive behavior. These objectives may include, for example, rehabilitation, emotion regulation, engagement, or difficulty adjustment. These objectives are pursued through computational \textit{adaptation methods}, such as rule-based logic, search algorithms, optimization procedures, or direct ML inference. Importantly, many of these methods overlap with those used in player experience modeling, which can obscure the boundary between them. Adaptation ultimately materializes in the form of \textit{adapted content}: the concrete elements of the game environment that are modified in response to model predictions. Examples include level layout, enemy behavior, task parameters, pacing, narrative events, interface elements, and audio-visual effects.

In this review, the distinction between affect-informed adaptation and affective adaptation is made at the level of the adaptation objective. Affect-informed adaptation refers to systems that use affective or experiential information while optimizing another objective, such as difficulty, engagement, performance, or rehabilitation progress. Affective adaptation refers to systems in which affective experience is itself an explicit adaptation objective. This distinction is central to the analysis of stress, anxiety, horror, and related states, because their presence in the loop does not necessarily mean that they function as adaptation objectives.

Although these levels describe the structural components of adaptation, many systems are also guided by broader design philosophies which provide a high-level rationale for how and why content should respond to changes in player state, often shaping the selection or interpretation of objectives, methods, and content types. Examples include challenge tuning through DDA~\cite{hunicke2005case}, dynamic game generation via Procedural Content Generation (PCG)~\cite{shaker2016procedural}, or affect-sensitive narrative structuring through orchestration~\cite{liapis2018orchestrating}. Although such strategies are not formalized as a separate analytical dimension in this review, they serve as an important conceptual backdrop that influences the overall adaptation loop.

\subsection{Identifying the Experience-Driven Loop}
\label{sec:concepts:idloop}
For this review, studies were treated as complete-loop systems only when player data informed a player experience model and the model output informed content adaptation during gameplay or across playable episodes (see Section~~\ref{sec:concepts:defloop}). Studies limited to sensing, offline modeling, static personalization, simulation-only adaptation, or conceptual architectures were treated as partial-loop work and excluded during screening. This boundary is both methodological and conceptual: the review focuses on empirically evaluated systems in which sensing, modeling, and adaptation are connected in an implemented game or game-based application.

\subsection{Focusing on Horror, Stress, and Anxiety}
\label{sec:concepts:hsa}

Although complete experience-driven systems can target many experiential outcomes~\cite{shaker2010towards,chanel2020user,makantasis2021pixels,yannakakis2010towards}, this review gives specific attention to horror, stress, and anxiety because they represent a focused affective problem space in which complete-loop adaptation is especially relevant. These states are important in high-arousal play~\cite{wijayarathna2022toward}, horror and suspense design~\cite{graja2020impact}, therapeutic applications~\cite{kritikos2021personalized}, training simulations, and other contexts where the regulation or induction of affective experience may be part of the intended design~\cite{takatalo2010presence, kaye2016exploring}.

The review does not assume that affective adaptation should be limited to these states. Instead, horror, stress, and anxiety are used as an analytical focus for assessing whether complete-loop systems treat affective state as an adaptation objective or use affective information only to support another objective, such as DDA, engagement, rehabilitation, or performance calibration.

\section{Methodology}
\label{sec:methods}

This systematic review was conducted and reported with reference to the PRISMA 2020 guidelines~\cite{page2021prisma}. The methodology describes the search strategy, databases, inclusion and exclusion criteria, screening procedure, extraction dimensions, and evidence-maturity indicator.

\subsection{Search Strategy}

Because this review spans game research, affective computing, PCG, rehabilitation~\cite{proencca2018serious,rodrigues2024participation,fonseca2025games}, mental health~\cite{emmelkamp2021virtual,fleming2017serious,fominykh2018conceptual}, and other applied domains, terminology varies across the relevant literature~\cite{rodrigues2024participation}. To balance coverage and specificity, the review used a targeted keyword strategy based on terms common in game research and related applied domains.

Preliminary search trials were conducted in October 2025 to refine the keyword strategy and assess its alignment with the inclusion criteria. The final database searches were conducted in February 2026 using the following keyword string:

\begin{quote}
(``Game'' OR ``Videogame'' OR ``Games'' OR ``Videogames'') AND (``Procedural Content Generation'' OR ``PCG'') AND (``Horror'' OR ``Stress'' OR ``Anxiety'') AND (``Adaptation'' OR ``Orchestration'' OR ``Experience-Driven'')
\end{quote}

Each keyword block corresponded to one dimension of the review scope. The game-related terms identified studies using games or videogames as the primary interactive medium. The PCG-related terms were used as high-precision indicators of content modification, since the review focuses on systems where player experience modeling is connected to adaptive changes in game content rather than only to affect recognition, player profiling, or post-hoc analysis. The horror-, stress-, and anxiety-related terms delimited the review to a focused affective problem space. Finally, the adaptation-related terms captured studies concerned with dynamic or personalized content adjustment, including work framed through experience-driven adaptation or orchestration~\cite{yannakakis2011experience,liapis2018orchestrating}.

The review considered publications from January 1, 2015 to December 31, 2025. This range was selected to focus on contemporary work following the emergence of experience-driven game research and the increasing use of affective and generative methods in adaptive systems.

\subsection{Inclusion and Exclusion Criteria}

Although the keyword strategy delimited retrieval around horror-related experiences, stress, and anxiety, eligibility was determined by whether each study implemented and empirically evaluated a complete experience-driven loop. After retrieval, studies were not excluded solely because their relation to horror, stress, or anxiety was indirect; this relation was treated as an analytical dimension of the selected studies, as described in Section~\ref{sec:methods:dims}. Eligibility was determined using the following inclusion and exclusion criteria:

\begin{itemize}
    \item Published in peer-reviewed journals or proceedings;
    \item Published between January~1,~2015 and December~31,~2025;
    \item Written in English;
    \item Experimental results include human players and report basic sample characteristics, at minimum the number of human players included;
    \item Player interaction resulting in game adaptation;
    \item Complete experience-driven loop.
\end{itemize}

The exclusion criteria were defined as follows:

\begin{itemize}
    \item Outside scope;
    \item Abstracts, front matter, and books;
    \item Conceptual models or theoretical analyses;
    \item Reviews or surveys;
    \item Written in languages other than English.
\end{itemize}

\subsection{Selected Databases} 
\label{sec:methods:dbs}
Searches were conducted in IEEE Xplore, ACM Digital Library, SpringerLink, Scopus, and b-on. These databases were selected to cover computer science, game research, affective computing, and applied health domains.

Due to the specific search interface and metadata structure of IEEE Xplore, the keyword string was adapted accordingly as follows:

\begin{quote}
(("Full Text \& Metadata":"Game" OR "Full Text \& Metadata":"Videogame" OR "Full Text \& Metadata":"Games" OR "Full Text \& Metadata":"Videogames") AND ("Full Text \& Metadata":"Procedural Content Generation" OR "Full Text \& Metadata":"PCG") AND ("Full Text \& Metadata":"Horror" OR "Full Text \& Metadata":"Stress" OR "Full Text \& Metadata":"Anxiety") AND ("Full Text \& Metadata":"Adaptation" OR "Full Text \& Metadata":"Orchestration" OR "Full Text \& Metadata":"Experience-Driven") )
\end{quote}

\subsection{Reviewing Procedures} 
\label{sec:methods:reviewprocs}

Search results were imported into Rayyan~\cite{ouzzani2016rayyan} for duplicate identification and screening management. Titles and abstracts were then independently screened by three reviewers. During this phase, systematic reviews, narrative reviews, surveys, vision papers, collections of abstracts or front matter, and studies from unrelated fields were excluded. Studies that failed to meet the inclusion criteria or met any exclusion criterion were also removed. Disagreements were resolved through discussion and consensus.

Records considered potentially relevant proceeded to full-text screening, where the same eligibility criteria were applied.

\subsection{Dimensions of Analysis}
\label{sec:methods:dims}

Following the PRISMA filtering process, each of the selected studies was systematically analyzed according to a set of predefined classification dimensions. These dimensions were designed to capture the methodological characteristics and design strategies employed in experience-driven adaptation systems. The classification scheme includes the following attributes:

\begin{itemize}

    \item \textbf{Sample size:} The number of participants involved in the study, used to assess the scale and empirical grounding of each implementation. This includes the reported total number of users and, where applicable, any division into control and experimental groups.

    \item \textbf{Input data:} Types of data used to model the player experience, as defined in Section~\ref{sec:concepts:defloop:pdata}.

    \item \textbf{Adaptation objectives:} The intended goals of the adaptation process, such as emotion regulation, engagement, or difficulty adjustment, discussed in Section~\ref{sec:concepts:defloop:contadapt}.
    \item \textbf{Modeling techniques:} Methods used to construct or update the player experience model, as discussed in Section~\ref{sec:concepts:defloop:pexpmodel}.

    \item \textbf{Adaptation methods:} The mechanisms through which content is modified in response to the experience model, such as rule-based logic, optimization strategies, search-based generation, or ML model inference (see Section~\ref{sec:concepts:defloop:contadapt}).

    \item \textbf{Game type:} The genre or nature of the game or application used in the study (e.g., platformer, role-playing game, horror, exergame, etc.).

    \item \textbf{Adapted content:} The specific game elements subjected to adaptation, as described in Section~\ref{sec:concepts:defloop:contadapt}.

    \item \textbf{Relation to stress, anxiety, or horror:} The extent to which the study explicitly addresses the affective states or experiences emphasized in this review, as framed in Section~\ref{sec:concepts:hsa}. This dimension distinguishes between studies where stress-, anxiety-, or horror-related states are absent or only contextual, studies where they appear as affective cues or secondary outcomes, and studies where they function as explicit adaptation targets. For presentation in Table~\ref{tab:papers_focus}, this relation was coded using a five-point visual scale: no or minimal explicit relation; indirect or contextual relation; partial affective relevance; strong explicit relevance; and direct use as an adaptation target.
\end{itemize}

This scheme enabled structured comparison across the selected studies and provided the basis for the synthesis reported in Section~\ref{sec:results}.

\subsection{Empirical Evidence Maturity}
\label{sec:methods:empirical}
To complement the descriptive classification, each selected study was assigned an evidence-maturity indicator. This indicator was not used as an inclusion or exclusion criterion and should not be interpreted as a formal risk of bias assessment. Instead, it provides a comparative indication of how directly and extensively each complete-loop system was empirically evaluated.

The evidence-maturity level for each study was coded by the authors through discussion and consensus. The coding considered the scale of the empirical evaluation, the presence of baseline or comparison conditions, the use of defined outcome measures, the extent of quantitative analysis, and whether the evaluation directly tested the effect of the adaptive mechanism rather than only demonstrating system feasibility. Limited evidence (L) refers to formative, exploratory, proof-of-concept, or small-scale evaluations, typically involving very small samples, descriptive analysis, informal feedback, or no baseline or comparison condition. Moderate evidence (M) refers to structured empirical evaluations with human participants and defined outcome measures, but with limitations such as small-to-moderate samples, partial comparison conditions, limited control over confounds, or incomplete evidence that the adaptive mechanism itself caused the observed effects. Stronger evidence (S) refers to comparatively mature evaluations, such as controlled studies, within- or between-subject comparisons, baseline or non-adaptive conditions, larger samples or deployment-scale testing, clear quantitative outcome measures, and analyses that directly assess the effect of adaptation.

The purpose of this coding was to distinguish studies that demonstrate the existence of a complete-loop system from studies that provide stronger empirical evidence about the effects of that system.

\section{Results}
\label{sec:results}
The systematic search and screening process yielded a final corpus of 23 studies that implemented and empirically evaluated the complete experience-driven loop. The resulting corpus was predominantly composed of difficulty-driven adaptive systems, with few studies targeting affective adaptation as a primary objective. This section first reports the PRISMA filtering process, then summarizes the main configurations of the selected corpus, and finally compares the studies across the extraction dimensions.

\subsection{PRISMA Filtering Results}
\label{sec:results:prisma}

\begin{figure}[t]
    \centering
    \resizebox{\columnwidth}{!}{
    \begin{tikzpicture}[node distance=7em and 5.5em]

    \definecolor{PrismaBlue}{HTML}{A2C3E3}
    \tikzset{
        base/.style={draw, },
        headers/.style={base, font=\bfseries, minimum height=1.8em, rounded corners, align=center},
        stage/.style={headers, anchor=north east, fill=PrismaBlue, rotate=90},
        top/.style={headers, anchor=north west, fill=Dandelion, minimum width=37.9em},
        prisbase/.style={base, anchor=north west, align=left},
        prisdata/.style={prisbase, minimum width=13.3em, minimum height=4.4em, text width=13.3em},
        prisinfo/.style={prisbase, minimum width=21.5em, text width=21.3em},
        arr/.style={->,>={Triangle[length=5pt,width=5pt]},rounded corners=10pt},
      }

    \node[top] (header) at (0, 0) {Identification of studies via databases and registers};

    \node[stage,minimum width=9.84em] (id) at (-0.9,-0.8) {Identification};
    \node[stage,minimum width=32.1em] (screen) at (-0.9, -4.8) {Screening};
    \node[stage] (included) at (-0.9, -16.6) {Included};

    \node[prisdata] (id_records) at (0, -0.8) {Records identified from:\\
      $\quad$Databases ($n=5$)\\
      $\qquad$ Springer ($118$)\\ 
      $\qquad$ IEEE ($54$)\\ 
      $\qquad$ ACM ($129$)\\ 
      $\qquad$ Scopus ($125$)\\ 
      $\qquad$ b-on ($246$)\\ 
      $\qquad$TOTAL: $672$};
    \node[prisinfo] (id_info) at (5.6, -0.8) {Records removed \textit{before screening}:\\
      $\quad$Duplicate records removed ($n=85$)}; 

    \node[prisdata] (scr_records1) at (0, -4.8) {Records screened\\($n=587$)\\\mbox{}}; 
    \node[prisinfo] (scr_info1) at (5.6, -4.8) {Records excluded ($n=415$)\\ 
      $\quad$Reason ``outside scope'' ($n=142$)\\
      $\quad$Reason ``abstract/front matter/book'' ($n=141$)\\ 
      $\quad$Reason ``conceptual model/theoretical analysis''\\
        $\qquad$($n=43$)\\
      $\quad$Reason ``reviews'' ($n=86$)\\ 
      $\quad$Reason ``not in English'' ($n=2$)\\
      $\quad$Reason ``duplicate'' ($n=1$)};

    \node[prisdata] (scr_records2) at (0, -9) {Reports sought for retrieval\\($n=172$)\\\mbox{}}; 
    \node[prisinfo] (scr_info2) at (5.6, -9) {Records excluded ($n=132$)\\ 
      $\quad$Reason ``outside scope'' ($n=53$)\\ 
      $\quad$Reason ``no human experiments'' ($n=18$)\\
      $\quad$Reason ``abstract/front matter/book'' ($n=2$)\\ 
      $\quad$Reason ``conceptual model/theoretical analysis''\\
        $\qquad$($n=10$)\\
      $\quad$Reason ``reviews'' ($n=13$)\\ 
      $\quad$Reason ``no loop/no adaptation'' ($n=34$)\\ 
      $\quad$Reason ``no results section'' ($n=2$)};

    \node[prisdata] (scr_records3) at (0, -13.4) {Reports assessed for eligibility\\($n=40$)\\\mbox{}}; 
    \node[prisinfo] (scr_info3) at (5.6, -13.4) {Records excluded ($n=17$)\\ 
      $\quad$Reason ``outside scope'' ($n=2$)\\
      $\quad$Reason ``no human experiments'' ($n=2$)\\ 
      $\quad$Reason ``conceptual model/theoretical analysis''\\
        $\qquad$($n=2$)\\
      $\quad$Reason ``no loop/no adaptation'' ($n=11$)}; 

    \node[prisdata] (incl_records) at (0, -16.6) {Studies included in review\\($n=23$)\\\mbox{}};

    \draw[arr] ($(id_records.north east)+(0, -1em)$)
    -- ($(id_info.north west)+(0, -1em)$);
    \draw[arr] ($(scr_records1.north east)+(0, -1em)$)
    -- ($(scr_info1.north west)+(0, -1em)$);
    \draw[arr] ($(scr_records2.north east)+(0, -1em)$)
    -- ($(scr_info2.north west)+(0, -1em)$);
    \draw[arr] ($(scr_records3.north east)+(0, -1em)$)
    -- ($(scr_info3.north west)+(0, -1em)$);

    \draw[arr] (id_records.south) -- (scr_records1.north);
    \draw[arr] (scr_records1.south) -- (scr_records2.north);
    \draw[arr] (scr_records2.south) -- (scr_records3.north);
    \draw[arr] (scr_records3.south) -- (incl_records.north);

    \end{tikzpicture}
    } 

    \caption{PRISMA 2020 flow diagram illustrating the systematic literature search and selection process. The figure outlines the number of records identified, screened, excluded, and ultimately included in the final review, based on the defined inclusion and exclusion criteria. A total of 23 studies met all conditions and were selected for in-depth analysis.}
    \label{fig:prisma}
\end{figure}
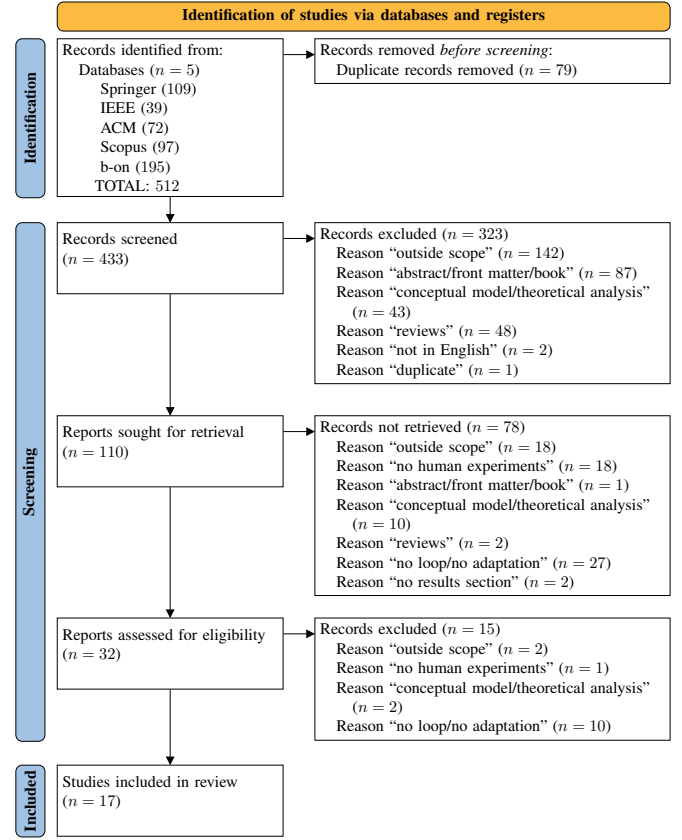

The PRISMA flowchart, shown in Fig.~\ref{fig:prisma}, provides a structured overview of the review process, illustrating the selection and screening of studies included in the analysis.

The systematic search identified 672 records across SpringerLink, IEEE Xplore, ACM Digital Library, Scopus, and b-on. After removing 85 duplicates, 587 records proceeded to title and abstract screening. This phase excluded 415 records, mainly because they were outside the review scope, were abstracts or front matter, presented conceptual or theoretical work, or were review publications.

Full-text retrieval and screening were conducted for the remaining records. At this stage, exclusions were mainly due to lack of human experiments, absence of a complete-loop implementation, absence of adaptive gameplay, or mismatch with the review scope. A total of 40 full-text articles were assessed for eligibility, of which 23 met all inclusion criteria and were included in the final corpus. The most consequential exclusion reason across later screening stages was the absence of a complete experience-driven loop, especially studies that addressed sensing, modeling, or content generation without connecting these components in an empirically evaluated adaptive game system.

Table~\ref{tab:papers} lists the 23 selected studies, including publication year, authors, and title. The following subsections summarize the main configurations of the selected corpus and then analyze the studies by extraction dimension.

\begin{table*}
\centering
\caption{Studies Included Following PRISMA Screening}
\begin{tabular}{lcll}
\toprule
Ref. & Year & Authors  & Title\\
\midrule

\cite{hocine2015adaptation}  & 2015 & Hocine et al. & \textit{Adaptation in serious games for upper-limb rehabilitation: an approach to improve training outcomes} \\

\cite{zook2015temporal}      & 2015 & Zook \& Riedl & \textit{Temporal game challenge tailoring}\\

\cite{nogueira2016vanishing} & 2016 & Nogueira et al. & \textit{Vanishing scares: biofeedback modulation of affective player experiences in a procedural horror game}\\

\cite{nagle2016towards}      & 2016 & Nagle et al. & \textit{Towards a system of customized video game mechanics based on player personality: relating the}\\
 & & & \textit{big five personality traits with difficulty adaptation in a first-person shooter game}\\

\cite{pirovano2016intelligent} & 2016 & Pirovano et al. & \textit{Intelligent game engine for rehabilitation (IGER)}\\

\cite{stein2018eeg}          & 2018 & Stein et al. & \textit{EEG-triggered dynamic difficulty adjustment for multiplayer games}\\

\cite{bicho2018multi}        & 2018 & Bicho \& Martinho & \textit{Multi-dimensional player skill progression modelling for procedural content generation}\\

\cite{mostefai2019generic}   & 2019 & Mostefai et al. & \textit{A generic and efficient emotion-driven approach toward personalized assessment and adaptation in} \\
& & & \textit{serious games}\\

\cite{blom2019modeling}      & 2019 & Blom et al. & \textit{Modeling and adjusting in-game difficulty based on facial expression analysis}\\

\cite{sifa2020matrix}        & 2020 & Sifa et al. & \textit{Matrix- and tensor factorization for game content recommendation}\\

\cite{pfau2020enemy}         & 2020 & Pfau et al. & \textit{Enemy within: long-term motivation effects of deep player behavior models for dynamic difficulty}\\
& & & \textit{adjustment}\\

\cite{lara2021induction}     & 2021 & Lara-Alvarez et al.  & \textit{Induction of emotional states in educational video games through a fuzzy control system}\\

\cite{huber2021dynamic}      & 2021 & Huber et al.  & \textit{Dynamic difficulty adjustment in virtual reality exergames through experience-driven procedural}\\
& & & \textit{content generation}\\

\cite{rosa2021dynamic}       & 2021 & Rosa et al. & \textit{Dynamic difficulty adjustment using performance and affective data in a platform game}\\

\cite{chen2023impact}        & 2023 & Chen et al. & \textit{Impact of BCI-informed visual effect adaptation in a walking simulator}\\

\cite{liang2023eeg}          & 2023 & Liang et al. & \textit{EEG-based VR scene adaptive generation system for regulating emotion}\\

\cite{fisher2024exploring}   & 2024 & Fisher \& Kulshreshth & \textit{Exploring dynamic difficulty adjustment methods for video games}\\

\cite{han2024exploring}      & 2024 & Han et al. & \textit{Exploring emotional responses with dynamic difficulty adjustment adaptation in immersive virtual}\\
& & & \textit{reality exergaming}\\

\cite{loizou2025framework}   & 2025 & Loizou \& Andreou & \textit{A framework for standardizing the development of serious games with real-time self-adaptation}\\
& & & \textit{capabilities using digital twins}\\

\cite{giariskanis2025dynamic}& 2025 & Giariskanis et al. & \textit{Dynamic difficulty adjustment in audio augmented reality games}\\

\cite{croissant2025advancing}& 2025 & Croissant et al. & \textit{Advancing methodological approaches in affect-adaptive video game design: empirical validation of}\\
& & & \textit{emotion-driven gameplay modification}\\

\cite{mario2025analyzing}    & 2025 & Mario \& Arifin & \textit{Analyzing the impacts of dynamic game balancing in a survival roguelike chess game}\\

\cite{shang2025general} & 2025 & Shang et al. & \textit{General purpose haptic/biometric-based dynamic difficulty adjustment for post-stroke upper-limb}\\
& & & \textit{rehabilitation games}\\

\bottomrule
\end{tabular}
\label{tab:papers}
\end{table*}

\subsection{Overview of the Selected Corpus}
\label{sec:results:papers}
The 23 selected studies are listed in Table~\ref{tab:papers} and characterized according to the extraction dimensions in Table~\ref{tab:papers_focus}. Rather than presenting the corpus only as a chronological sequence of individual studies, this subsection summarizes the main configurations of complete experience-driven adaptation observed in the selected papers. These configurations are not mutually exclusive; several studies combine difficulty-oriented adaptation with affective, physiological, or applied-domain components. Study-level dimensions are analyzed in Subsection~\ref{sec:results:papermethods}, while detailed summaries of all included studies are provided in Appendix~\ref{app:detailed-summaries}.

\textit{Difficulty- and performance-oriented adaptation.}
A first and dominant configuration uses the experience-driven loop primarily for difficulty, performance, or task calibration. In these systems, player data is used to estimate ability, progress, performance, or challenge, and the resulting model informs adaptations to tasks, enemies, pacing, level structure, or player parameters. This pattern appears across entertainment games, such as role-playing games, platformers, shooters, and roguelike prototypes~\cite{zook2015temporal,nagle2016towards,bicho2018multi,sifa2020matrix,pfau2020enemy,mario2025analyzing}, as well as applied contexts such as rehabilitation, educational games, and exergames~\cite{hocine2015adaptation,pirovano2016intelligent,huber2021dynamic,han2024exploring,loizou2025framework,giariskanis2025dynamic}. Although these systems differ in genre and technical implementation, they commonly treat adaptation as a means of maintaining an appropriate level of challenge, sustaining engagement, supporting training, or improving task performance.

\textit{Affect-informed adaptation.}
A second configuration incorporates affective, physiological, or expressive information into the loop, but still uses this information largely in service of difficulty or engagement-related adaptation. Examples include EEG-triggered or emotion-informed DDA, facial-expression-based difficulty adjustment, EDA-informed platform adaptation, and biometric rehabilitation systems~\cite{stein2018eeg,blom2019modeling,rosa2021dynamic,fisher2024exploring,shang2025general}. These studies show that affective information can be operationally integrated into adaptive game systems. However, they also illustrate the distinction central to this review: affective or physiological cues may contribute to the player experience model without affective state itself becoming the primary adaptation target.

\textit{Affective adaptation.}
A smaller subset of studies treats affective experience as an explicit adaptation objective. Nogueira et al.~\cite{nogueira2016vanishing} most closely matches the affective adaptation focus of this review, combining physiological modeling of arousal and valence with adaptations to layout, audiovisual effects, enemy behavior, and player attributes in a procedural horror game. Other studies address emotion regulation, emotional state induction, or affective gameplay modification in educational games, VR environments, or arcade-style games~\cite{mostefai2019generic,lara2021induction,liang2023eeg,croissant2025advancing}. These works demonstrate that complete affective adaptation is possible, but they remain comparatively rare within the selected corpus.

\textit{Applied and serious-game contexts.}
The corpus also shows that complete-loop adaptation is frequently explored in applied or serious-game contexts. Rehabilitation, education, training, and health-related systems appear repeatedly, often because these domains provide clear adaptation objectives and practical reasons for personalization~\cite{hocine2015adaptation,pirovano2016intelligent,huber2021dynamic,loizou2025framework,shang2025general}. However, these studies also tend to involve small samples, constrained interaction settings, or domain-specific adaptation criteria. By contrast, large-scale deployments are rare, with Sifa et al.~\cite{sifa2020matrix} standing out because of its commercial online-game setting and unusually large participant base.

Overall, the selected corpus shows that complete experience-driven loops have been implemented across several genres and application domains. However, most implementations remain oriented toward DDA, engagement, performance, or rehabilitation outcomes. Explicit affective adaptation, especially adaptation targeting stress, anxiety, horror, or related high-arousal experiences, appears only in a limited number of studies. This overview provides the basis for the comparative analysis that follows, where these patterns are examined across input data, adaptation objectives, modeling techniques, adaptation methods, game types, adapted content, and relation to stress, anxiety, or horror.

\begin{table*}[!t]
\centering
\caption{Classification of the selected complete-loop studies according to the extraction dimensions defined in Section~\ref{sec:methods:dims}.}
\begin{tabular}{crllllllll}
\toprule
     & Sample & Input & Adaptation & Modeling   & Adaptation   & Game & Adapted & H/S/A & Evid.\\
Ref. & Size   & Data  & Objectives  & Techniques & Methods      & Type & Content & Related & Mat.\\
\midrule

\arrayrulecolor{black!20}
\cite{hocine2015adaptation} & 7 & PI, GT & Rehab., & Bio-inspired & MCTS & Platformer & Layout,  & {\tiny\LEFTcircle\Circle\Circle\Circle\Circle} & M \\
{\tiny\color{gray}Hoci.} & & & DDA & & & & incentives & & \\

\hdashline[1pt/1pt]

\cite{zook2015temporal} & 32, 30 & GT & DDA & TF & ASP & RPG & Enemy selection  & {\tiny\CIRCLE\Circle\Circle\Circle\Circle} & M \\
{\tiny\color{gray}Zook} & & & & & & (turn-based) &  & & \\

\hdashline[1pt/1pt]

\cite{nogueira2016vanishing} & 24 & GT, PH & Emotion & Regression, & Rule-based & Survival  & Layout, SFX, & {\tiny\CIRCLE\CIRCLE\CIRCLE\CIRCLE\CIRCLE} & S \\
{\tiny\color{gray}Nogu.} & & &  & rule-based & & horror & VFX, enemy attr., & & \\
 & & & & & & & player attr. & &\\

\hdashline[1pt/1pt]

\cite{nagle2016towards} & 160, 80 & GT, QT & Engage., & Linear regression & Rule-based & FPS & Enemy attr. & {\tiny\CIRCLE\Circle\Circle\Circle\Circle} & S \\
{\tiny\color{gray}Nagle} & & & DDA & & & & & & \\

\hdashline[1pt/1pt]

\cite{pirovano2016intelligent} & 7 & BE, GT, & Rehab., & Fuzzy-based & Bayesian & Exergame & Exercise attr. & {\tiny\CIRCLE\Circle\Circle\Circle\Circle} & L \\
{\tiny\color{gray}Pirov.} & & TPA & DDA & evaluation &  optimization & & & & \\

\hdashline[1pt/1pt]

\cite{stein2018eeg} & 24 & GT, PH & Engage., & Proprietary, & Rule-based & TPS & Player attr., & {\tiny\LEFTcircle\Circle\Circle\Circle\Circle} & S \\
{\tiny\color{gray}Stein}& &  & DDA & thresholding  & & (multiplayer)  & enemy attr., & & \\
& &  &  & & & & pacing  & & \\

\hdashline[1pt/1pt]

\cite{bicho2018multi} & 30 & GT & Engage., & Rolling window, & Rule-based, & Platformer & Layout,  & {\tiny\CIRCLE\Circle\Circle\Circle\Circle} & M \\
{\tiny\color{gray}Bicho} & &   & DDA & A/B testing & log. scaling &  & incentives & & \\

\hdashline[1pt/1pt]

\cite{mostefai2019generic} & 30 & GT, QT & Emotion, & Rule-based & Markov chains, & Educational  & Tasks, pacing & {\tiny\CIRCLE\Circle\Circle\Circle\Circle} & S \\
{\tiny\color{gray}Moste.} & & & DDA & & rule-based & game (prog.)&  & &\\

\hdashline[1pt/1pt]

\cite{blom2019modeling} & 10, 25 & BE, GT, & Emotion, & Proprietary, & GAO, rule-based, & Platformer & Layout & {\tiny\CIRCLE\LEFTCIRCLE\Circle\Circle\Circle} & S \\
{\tiny\color{gray}Blom} & & QT & DDA  & RFC (pre-trained) & RFC (inference) &  & & & \\

\hdashline[1pt/1pt]

\cite{sifa2020matrix} & 25$\,$686 & GT & Retention, & TF, MF, & TF/MF inference, & RPG & Task & {\tiny\LEFTcircle\Circle\Circle\Circle\Circle} & S \\
{\tiny\color{gray}Sifa}& & & DDA  & rule-based, & rule-based, &  & recommend. & & \\
& & & & neigh.-based & neigh.-based &  &  & & \\

\hdashline[1pt/1pt]

\cite{pfau2020enemy} & 171 & GT & Motivation, & Rule-based, & Rule-based, & MMORPG & Enemy attr.  & {\tiny\CIRCLE\LEFTcircle\Circle\Circle\Circle} & S \\
{\tiny\color{gray}Pfau}& & & DDA & MLP (train) & MLP (inference) & (solo runs)  & & & \\

\hdashline[1pt/1pt]

\cite{lara2021induction} & 40 & GT, BE  & Emotion, & Rule-based, SVR & Fuzzy control, & Educational  & SFX, tasks & {\tiny\CIRCLE\CIRCLE\CIRCLE\LEFTcircle\Circle} & S \\
{\tiny\color{gray}Lara}& & & DDA & & rule-based & game (math) & & & \\

\hdashline[1pt/1pt]

\cite{huber2021dynamic} & 19 & QT & Engage.,  & DRL & DRL (inference)& Exergame, & Layout & {\tiny\LEFTcircle\Circle\Circle\Circle\Circle} & L \\
{\tiny\color{gray}Huber}& &  & DDA &  (part. pre-trained) & & VR &  & &\\

\hdashline[1pt/1pt]

\cite{rosa2021dynamic} & 20, 36, & GT, PH & Engage., & Rule-based & Rule-based & Platformer & Layout, & {\tiny\CIRCLE\Circle\Circle\Circle\Circle} & S \\
{\tiny\color{gray}Rosa}& 132 & & DDA & & & & player attr. & &\\

\hdashline[1pt/1pt]

\cite{chen2023impact} & 24 & GT, PH & Immersion & Rolling window, & Rule-based & Walking sim. & VFX & {\tiny\LEFTcircle\Circle\Circle\Circle\Circle} & M \\
{\tiny\color{gray}Chen}& & & & rule-based & & & & & \\

\hdashline[1pt/1pt]

\cite{liang2023eeg} & 80 & PH & Emotion & Proprietary & Rule-based & Walking sim., & Decorations & {\tiny\CIRCLE\CIRCLE\CIRCLE\CIRCLE\Circle} & M \\
{\tiny\color{gray}Liang} & & & &  & & VR & & &\\

\hdashline[1pt/1pt]

\cite{fisher2024exploring} & 31 & GT, PH & Emotion, & Proprietary, & Rule-based & FPS &  Player attr., & {\tiny\CIRCLE\CIRCLE\Circle\Circle\Circle} & S \\
{\tiny\color{gray}Fisher} & & & DDA &  rule-based, & & (survival) & enemy attr., & & \\
 & & & & moving avg. & & &  pacing & &\\

\hdashline[1pt/1pt]

\cite{han2024exploring} & 14 & GT & DDA & Rule-based & Rule-based & Exergame, & Pacing & {\tiny\CIRCLE\Circle\Circle\Circle\Circle} & M \\
{\tiny\color{gray}Han} & & &   &   & &  VR & & &\\

\hdashline[1pt/1pt]

\cite{loizou2025framework} & 5 & GT & Engage., & Rule-based & Rule-based & Speech therapy &
Task attr.,  &
{\tiny\LEFTcircle\Circle\Circle\Circle\Circle} & L \\
{\tiny\color{gray}Loizou} & & & DDA &  & &  game & pacing, hints, & &\\
 & & & &  & &  & visuals, sounds & &\\

\hdashline[1pt/1pt]

\cite{giariskanis2025dynamic} & 20 & GT & Engage., & Rule-based & Rule-based & Edu. game & Task attr., & {\tiny\CIRCLE\Circle\Circle\Circle\Circle} & L \\
{\tiny\color{gray}Giari.} & & & DDA & &  & (culture), AR & pacing & & \\

\hdashline[1pt/1pt]

\cite{croissant2025advancing}& 161, 158 & GT, QT & Emotion, & Linear regression & Statistical, & 2D action & Enemy attr. & {\tiny\LEFTcircle\Circle\Circle\Circle\Circle} & S \\

{\tiny\color{gray}Crois.} & & & DDA &  & rule-based  & arcade & & &\\

\hdashline[1pt/1pt]

\cite{mario2025analyzing} & 14 & GT & Engage., & Linear & Rule-based & Roguelike & Enemy attr., & {\tiny\LEFTcircle\Circle\Circle\Circle\Circle} & L \\
{\tiny\color{gray}Mario} & & & DDA & combination &  & & task attr., hints, & &\\
 & & & & &  &  & visuals, pacing & & \\

\hdashline[1pt/1pt]

\cite{shang2025general} & 11 & PI, GT, & Rehab., & Belief model & Rule-based & Exergame & Task attr., hints,  & {\tiny\CIRCLE\CIRCLE\Circle\Circle\Circle} & L \\
{\tiny\color{gray}Shang} & & PH & DDA & & & & haptic feedback, & & \\
& & & & &  &  & visuals, audio & & \\

\arrayrulecolor{black}
\bottomrule
\multicolumn{10}{p{1.84\columnwidth}}{
\textit{Note.} The H/S/A (horror, stress, and/or anxiety) column uses the five-point relation coding defined in Section~\ref{sec:methods:dims}. Evid. Mat. = evidence maturity (Section~\ref{sec:methods:empirical}): L = limited, M = moderate, S = stronger.
}\\
\multicolumn{10}{p{1.84\columnwidth}}{
PI - Peripheral input; GT - Game telemetry; PH - Physiology; QT - Questionnaires; TPA - Third-person annotation; BE - Bodily expressions; DDA - Dynamic difficulty adjustment; RFC - Random forest classifier; MCTS - Monte-Carlo tree search; TF - Tensor factorization; MF - Matrix factorization; MLP - Multi-layer perceptron; SVR - Support vector regression; DRL - Deep reinforcement learning; ASP - Answer set programming; GAO - Gradient ascent optimization; RPG - Role-playing game; FPS - First-person shooter; TPS - Third-person shooter; MMORPG - Massive multiplayer online RPG; VR - Virtual reality; AR - Augmented reality; NPC - Non-player character.
}\\
\end{tabular}
\label{tab:papers_focus}
\end{table*}

\subsection{Comparative Analysis by Dimension}
\label{sec:results:papermethods}

Table~\ref{tab:papers_focus} presents the 23 selected studies according to the dimensions defined in Section~\ref{sec:methods:dims}. Most studies used small-to-moderate samples, commonly in the 20--40 participant range, with notable exceptions. Several applied or rehabilitation-oriented studies involved very small samples, reflecting recruitment and intervention constraints in these contexts~\cite{hocine2015adaptation,pirovano2016intelligent,loizou2025framework,shang2025general}. At the other end of the scale, Sifa et al.~\cite{sifa2020matrix} analyzed more than 25$\,$000 participants in a commercial online game setting.

\begin{figure}
  \centering
  \includegraphics[width=1\columnwidth]{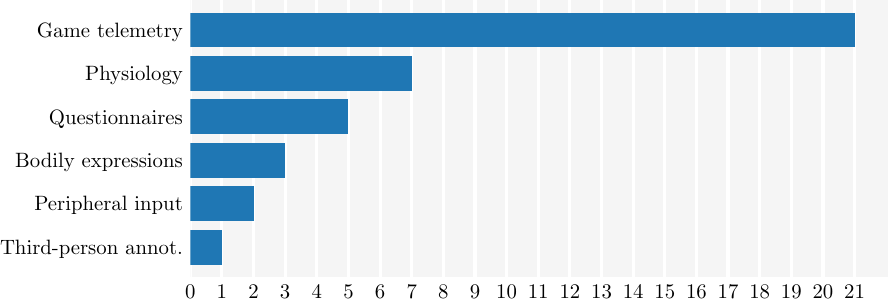}
  \caption{Frequency of input data types in the selected studies. \textit{Game telemetry} includes both gameplay (player metrics) and game context (instant game state) data. For this review, \textit{Physiology} consists of electrodermal activity (EDA), heart rate (HR), electromyography (EMG), electroencephalography (EEG), and functional near-infrared spectroscopy (fNIRS) signals, while \textit{Bodily expressions} refer to speech, facial behavior, and motion ranges.}
  \label{fig:stats:inputdata}
\end{figure}

In terms of input data, 13 of the 23 selected studies used more than one input type. Game telemetry (GT), including gameplay and game-context data, was the predominant source (Fig.~\ref{fig:stats:inputdata}). In most cases, telemetry was used as gameplay data, such as performance measures, death locations, jump success rates, kill ratios, or distance between players~\cite{rosa2021dynamic,stein2018eeg}. Chen et al.~\cite{chen2023impact} was an exception, using game-context information in the non-BCI version of the experiment.

Physiological data and questionnaires were also used in several studies, with EEG, EDA, and HR being the most common physiological measures~\cite{nogueira2016vanishing,stein2018eeg,liang2023eeg,rosa2021dynamic,fisher2024exploring,shang2025general}. By contrast, bodily expressions and peripheral input appeared in only five studies in total. Given prior work on affect inference from interaction behavior and common input devices~\cite{yang2021review,bahreini2016data,cowie2001emotion}, these modalities remain a relevant but less represented direction within the selected corpus.

\begin{figure}
  \centering
  \includegraphics[width=1\columnwidth]{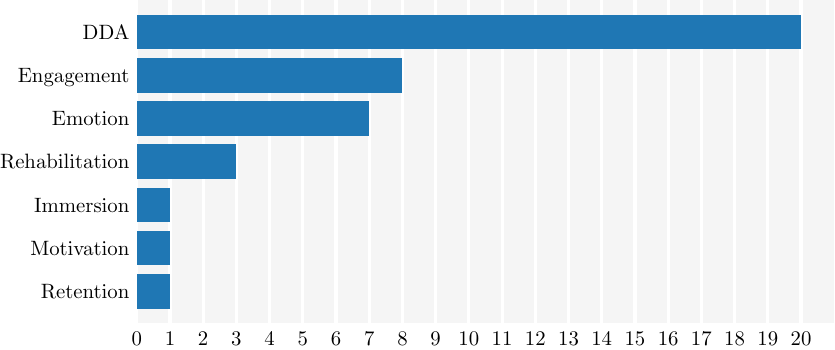}
  \caption{Frequency of self-reported adaptation objectives in the selected studies, with each study mentioning at most two clear objectives.}
  \label{fig:stats:adaptobj}
\end{figure}

A quantitative summary of adaptation objectives in the selected studies is depicted in Fig.~\ref{fig:stats:adaptobj}. For each study, one or two main adaptation objectives were carefully extracted. While some of these are closely related, in particular engagement, retention, motivation, and immersion, the selected studies were generally clear about the targeted adaptation objectives. The most common adaptation objective across the selected studies is DDA (20 studies), typically in combination with other goals; the only exceptions were the works of Zook \& Riedl~\cite{zook2015temporal} and Han et al.~\cite{han2024exploring}, which focus exclusively on DDA. GT input data was used in 19 of the 20 DDA-focused studies, underlining the reliance of DDA on gameplay and game context data. Huber's VR exergame study~\cite{huber2021dynamic} is the exception, relying only on post-level player feedback to adjust difficulty.

Following DDA, engagement was the adaptation objective in eight studies, followed by emotion in seven studies (the category most directly aligned with the affective focus of this review, as discussed at the end of this section) and rehabilitation in three. Although DDA was not the primary motivation for this review, it emerged as the dominant adaptation objective in the selected corpus.

Importantly, the presence of affective information did not necessarily correspond to affective adaptation as an objective. Of the 20 studies coded as DDA-oriented, six were classified as limited evidence, four as moderate evidence, and ten as stronger evidence. This indicates variation in evidence maturity, but also shows that complete-loop evaluation activity in the selected corpus remains organized primarily around difficulty, challenge, engagement, rehabilitation, or performance calibration. A similar pattern appears in the emotion-oriented subset: five of the seven studies coded with emotion as an adaptation objective also included DDA. These results reinforce the central distinction of this review: affective or emotion-related information is often incorporated into complete-loop systems whose adaptive logic remains tied to challenge or performance regulation rather than to affective state as an independent adaptation target.

\begin{figure}
  \centering
  \includegraphics[width=1\columnwidth]{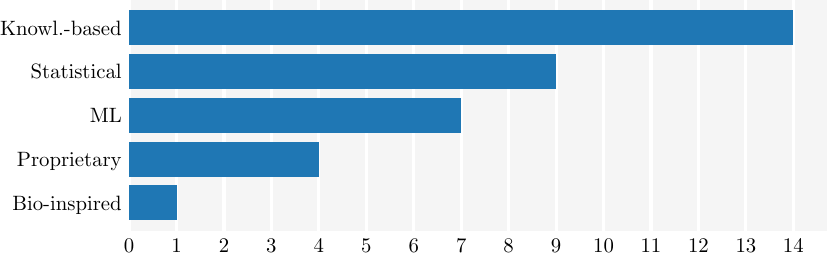}
  \caption{Frequency of player experience modeling techniques used in the selected studies. Rule-based, fuzzy methods, and belief models were incorporated into \textit{Knowledge-based} approaches. \textit{Statistical} techniques feature regression, thresholding, rolling window, moving average, A/B testing, and neighborhood-based methods. \textit{Machine Learning (ML)} techniques encompass tensor and matrix factorization, random forest classifiers, support vector regression, and neural network-based methods.}
  \label{fig:stats:modeltec}
\end{figure}

\begin{figure}
  \centering
  \includegraphics[width=1\columnwidth]{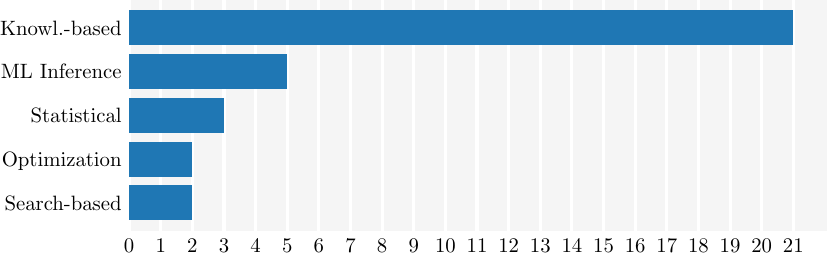}
  \caption{Frequency of content adaptation methods used in the selected studies. Here, the \textit{Knowledge-based} category includes rule-based approaches, fuzzy methods, and logarithmic scaling. \textit{ML Inference} consists of directly inferenced machine learning (ML) models, namely latent factor models from tensor factorization and matrix factorization, random forest classifiers, and neural network-based models. \textit{Statistics} includes Markov chains, distance-based models, and linear regression. Bayesian optimization and gradient ascent optimization are categorized as \textit{Optimization}. \textit{Search-based} approaches consist of Monte Carlo Tree Search and Answer Set Programming.}
  \label{fig:stats:adaptmet}
\end{figure}

Several selected studies did not clearly separate player experience modeling from content adaptation, so the distinction between modeling techniques and adaptation methods was coded on a best-effort basis. Modeling techniques were treated as methods that transform input data into a representation of player state or experience, whereas adaptation methods were treated as mechanisms that use that representation to select or modify game content. Because some studies used more than one technique, the frequencies in Figs.~\ref{fig:stats:modeltec} and~\ref{fig:stats:adaptmet} may include multiple entries per study. Rule-based, fuzzy, heuristic, and theory-driven approaches were grouped as knowledge-based methods.

Looking solely at modeling techniques, Fig.~\ref{fig:stats:modeltec} shows that knowledge-based approaches dominate, with 14 instances, followed by statistical techniques with 9 and ML methods with 7. Despite the broader growth of AI-driven approaches, the selected studies more often used rule-based, fuzzy, or statistical methods than fully data-driven ML pipelines. Four studies used proprietary algorithms for player experience modeling, typically embedded in commercial sensing or affect-recognition hardware~\cite{stein2018eeg,blom2019modeling,liang2023eeg,fisher2024exploring}. These tools can provide ready-to-use affective outputs, but they may also limit reproducibility because the underlying algorithms are closed, hardware availability can change, and long-term software support is uncertain.

With respect to content adaptation methods, Fig.~\ref{fig:stats:adaptmet} again shows a strong preference for knowledge-based approaches, with 21 use cases in the selected studies. ML inference appears less often, with five instances, followed by statistical, optimization, and search-based methods. Not all ML-based player experience models were directly used as adaptation methods; for example, Zook and Riedl~\cite{zook2015temporal} used tensor factorization for player modeling and answer set programming for content adaptation.

\begin{figure}
  \centering
  \includegraphics[width=1\columnwidth]{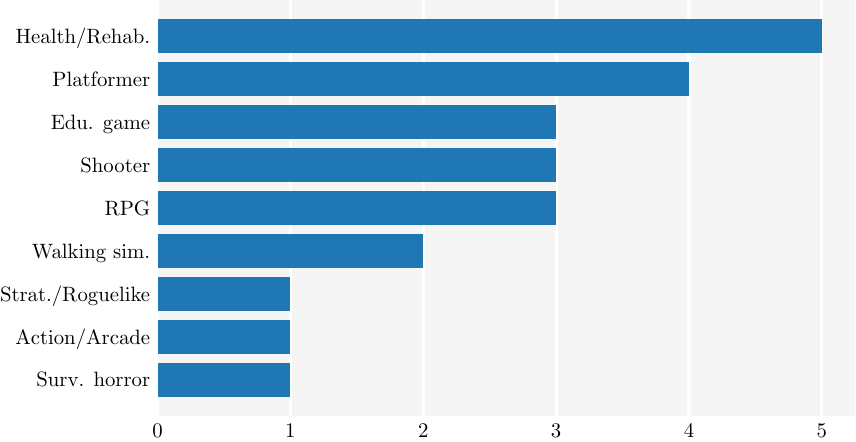}
  \caption{Frequency of game types used in the selected studies. First-person and third-person shooter games are grouped under the \textit{Shooter} category.}
  \label{fig:stats:gametype}
\end{figure}
\begin{figure}
  \centering
  \includegraphics[width=1\columnwidth]{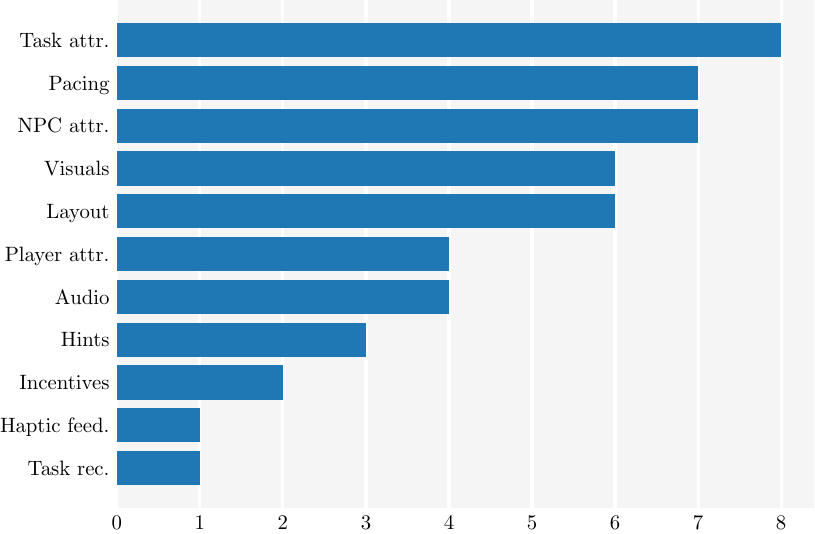}
  \caption{Frequency of adapted content in the selected studies. The \textit{Task attributes} category contains enemy selection and placement, exercise attributes, and tasks, while enemy attributes are grouped under \textit{NPC attributes}.}
  \label{fig:stats:adaptcont}
\end{figure}

Game types and adapted content were more varied (Figs.~\ref{fig:stats:gametype} and~\ref{fig:stats:adaptcont}). Health and rehabilitation games were the most common category, followed by platformers, educational games, shooters, and RPGs. Adaptation most often concerned task attributes, pacing, NPC attributes, visuals, and layout. These patterns support the applied orientation of many complete-loop systems, but they are secondary to the central finding that most systems adapted content for DDA, engagement, rehabilitation, or performance-related objectives rather than for affective state as the primary target.

Finally, the H/S/A relation coding shows that most selected complete-loop systems had limited explicit relation to stress, anxiety, or horror as adaptation objectives, even though retrieval was delimited around this affective problem space. This pattern supports the main interpretation of the review: affective states may appear in the motivation, input data, or evaluation of adaptive game systems without becoming the explicit objective of content adaptation.

\section{Discussion}
\label{sec:discussion}

The selected corpus shows that complete-loop adaptation is implemented in practice, but that most implementations use affective or experiential information to support difficulty, engagement, rehabilitation, or performance-related objectives rather than to target affective state directly. This section interprets that pattern, discusses the implementation difficulty of complete experience-driven loops, answers the research questions, and outlines the limitations and implications.

\subsection{Main Technical Findings}
\label{sec:discussion:techs}
The main technical pattern in the selected corpus is the dominance of DDA and related performance-oriented objectives. These systems most often relied on knowledge-based methods, including rule-based and fuzzy approaches, because they are interpretable, require less training data, and can be deployed in constrained settings. ML methods appeared in several studies, but were less frequent and often limited by data requirements, transparency, and runtime integration challenges. RL remains rare within the selected corpus. Only Huber et al.~\cite{huber2021dynamic} used a DRL-based adaptation approach, suggesting that interaction-based learning remains underrepresented among empirically evaluated complete-loop systems. Outside this corpus, approaches such as the EDPCGRL arachnophobia exposure therapy of Mahmoudi-Nejad et al.~\cite{mahmoudi2021arachnophobia} illustrate the potential of RL for affect-sensitive adaptation, although its evaluation with virtual/artificial subjects rather than human players placed it beyond our inclusion criteria.

Game telemetry emerged as the dominant input modality, most likely due to its ease of collection, non-intrusiveness, and suitability for large-scale studies, especially when compared to more intrusive data sources such as physiological signals or questionnaires. Within the selected corpus, less intrusive input methods, such as facial expression analysis or peripheral input (e.g., keyboard and mouse), were less common than game telemetry and physiology, despite their accessibility and ecological validity~\cite{yang2021review, baltruvsaitis2016openface}.

Game type and adapted-content distributions further show that complete-loop systems are often developed in applied contexts, especially health, rehabilitation, education, and training. However, these dimensions are secondary to the central pattern: the adapted content usually served difficulty, engagement, rehabilitation, or performance-related objectives rather than affective state as the primary target.

The evidence-maturity coding further qualifies these findings. Several studies connected player data, player experience modeling, and adaptive content, but evaluated the resulting systems mainly through feasibility, usability, descriptive feedback, or small-scale demonstrations. These studies are valuable because they show that complete-loop systems can be implemented in constrained settings, but they provide more limited evidence about whether the adaptive mechanism itself produced the intended experiential or affective effect.

\subsection{Affective Information Does Not Necessarily Imply Affective Adaptation}
\label{sec:discussion:reviewgoals}
This review was motivated by adaptive systems that explicitly address affective experiences related to horror, stress, and anxiety within complete experience-driven loops. The findings reveal that such systems remain scarce. Although several selected studies incorporate affective, physiological, or experiential information, this information often supports difficulty adjustment, engagement maintenance, rehabilitation progress, or performance calibration rather than serving as the basis for adaptation objectives in their own right.

This distinction is important for interpreting the corpus. The dominance of DDA does not mean that affective computing is absent from adaptive game systems. Rather, it suggests that affective information is frequently operationalized through difficulty regulation. Such systems may still be valuable, especially when challenge calibration is linked to engagement, flow, learning, or therapeutic progress. However, they differ from systems that explicitly aim to induce, regulate, sustain, or attenuate affective states such as stress, anxiety, fear, or tension.

Within the selected corpus, Nogueira et al.~\cite{nogueira2016vanishing} most closely matches the affective adaptation focus of this review, as it combines physiological modeling with horror-related affective modulation across multiple game-content dimensions. Other studies, such as Lara-Alvarez et al.~\cite{lara2021induction}, Liang et al.~\cite{liang2023eeg}, and Fisher and Kulshreshth~\cite{fisher2024exploring}, demonstrate related, but less complete, forms of stress- or emotion-oriented adaptation. Overall, however, the reviewed studies point to a gap between affective sensing and affective intervention: affective states may be measured, inferred, or used as auxiliary information, but are rarely the explicit objective around which the complete adaptation loop is organized.

\subsection{Interpretation and Implementation of the Experience-Driven Loop}
\label{sec:discussion:edloop}

Within the screened literature, 23 studies met the criteria for fully implementing and empirically evaluating the complete experience-driven loop by integrating its three foundational components: data acquisition, predictive modeling of player experience, and dynamic adaptation of game content. These studies captured affective and/or behavioral signals and also translated them into timely and context-sensitive adaptations. For example, the studies by Nogueira et al.~\cite{nogueira2016vanishing} and Chen et al.~\cite{chen2023impact} exemplify this complete-loop structure. Both works demonstrate continuous sensing, the use of internal models to interpret player state, and systematic content adjustment during gameplay, aligning closely with the complete-loop architecture defined in Section~\ref{sec:concepts}. These implementations highlight the feasibility of operationalizing the experience-driven loop in real-world game systems.

The screened and excluded records suggest fragmentation across sensing, modeling, and adaptation work. Despite extensive academic discussion of the experience-driven loop~\cite{yannakakis2011experience}, most studies have addressed only isolated components, such as affect recognition or PCG, without completing the loop via player experience models and adaptive feedback. This fragmentation highlights an ongoing challenge: translating affective and behavioral signals into meaningful, responsive gameplay that evolves continuously with the player's experience.

These records also illustrate why the complete-loop criterion was restrictive. Common partial-loop patterns included static or one-time personalization without ongoing adaptation~\cite{palma2021player,ince2021bilstm}, affect recognition or parameter analysis without implemented content adaptation~\cite{lopes2017modelling,graja2020impact,campo2023statistical,nogueira2015modelling,bevilacqua_game-calibrated_2019,orozco-mora_stress_2022}, architectures or prototypes that did not empirically close the loop~\cite{badia2019toward,frommel2018towards}, simulation-only evaluation~\cite{reis2023automatic}, and PCG work without real-time player experience modeling~\cite{acevedo2022procedural,ripamonti2017procedural}. These examples support the need for clearer reporting of sensing, player experience modeling, and content adaptation components.

This separation appears especially relevant for stress and anxiety, where several studies model affective responses to game stimuli or content changes without using those models to drive adaptive content~\cite{lopes2017modelling,graja2020impact}. Within this review's scope, the scarcity of complete experience-driven loops appears to reflect both limited integration across sensing, modeling, and adaptation work and the practical difficulty of coordinating these components in real time. Compared with many DDA implementations, which often map one performance metric to one difficulty parameter, affective adaptation may require multiple sensing modalities, ambiguous state inference, and coordinated changes across several game dimensions~\cite{liapis2018orchestrating}.

Among the selected studies, only the work of Nogueira et al.~\cite{nogueira2016vanishing}, highlighted earlier as an exemplary complete-loop implementation, approaches this level of multi-dimensional orchestration by adapting level layout, audiovisual effects, enemy attributes, and player attributes from a physiological model of arousal and valence. While other selected studies also adapt multiple content dimensions (e.g., \cite{loizou2025framework,shang2025general}), these do so primarily through performance-based difficulty adjustments applied across several game parameters, rather than through an integrated affective model driving coordinated content changes. Within this review scope, the rarity of such implementations appears to reflect the difficulty of integrating these components into a cohesive and empirically validated pipeline.

\subsection{Answering the Research Questions}
\label{sec:discussion:rqs}

For RQ1, this review found that complete-loop implementations were relatively uncommon within the retrieved corpus. After screening 672 records, 23 studies met all inclusion criteria and implemented player data acquisition, player experience modeling, and content adaptation in an empirically evaluated game or game-based system.

For RQ2, the selected complete-loop systems were predominantly oriented toward DDA, engagement, rehabilitation, or performance-related objectives. Game telemetry was the dominant input modality, and knowledge-based methods, including rule-based and fuzzy approaches, were the most common modeling and adaptation approaches. ML methods appeared in several studies but were less frequent.

For RQ3, stress, anxiety, and closely related affective states were rarely treated as explicit adaptation objectives within the selected complete-loop systems. In several systems, affective, physiological, or experiential cues appeared in the loop, but they more often supported challenge calibration, engagement, rehabilitation progress, or performance regulation than affective state as an independent adaptation objective.

\subsection{Limitations of this Review}
\label{sec:discussion:limitations}

This review faced several limitations. First, the keyword strategy was targeted rather than exhaustive of all affect-adaptive or biofeedback-based game research. Its purpose was to identify empirically evaluated systems in which player-state modeling is connected to concrete game-content adaptation, with particular emphasis on stress-, anxiety-, and horror-related affective experience. Consequently, studies framed through alternative terminology, such as biofeedback, affective interaction, or physiological self-regulation, may be underrepresented if they do not explicitly connect player-state modeling to content generation, orchestration, or adaptation. This issue is compounded by terminological fragmentation across communities: work in clinical, rehabilitation, and training contexts is often presented as a platform, simulator, or intervention rather than as a game~\cite{rodrigues2024participation}, which can make game-based adaptive systems difficult to retrieve through game-centered search terms. These factors delimit the scope of the review and should be considered when interpreting the findings. For example, Bian et al.~\cite{bian2019design} use physiological signals to adapt task difficulty in a VR driving system implemented with game technology, but primarily framed as a clinical driving-skill intervention. This borderline case is consistent with the pattern identified here: affective and physiological information may support task calibration without making affective state itself the adaptation objective.

Second, although this review introduced a compact evidence-maturity coding to contextualize how extensively and directly each adaptive system was empirically evaluated, this appraisal should not be interpreted as a formal assessment of methodological quality, risk of bias, or construct validity. For example, several included studies rely on physiological or affective signal interpretations whose robustness was not independently re-evaluated in this review, including studies using EDA, HR, EEG, facial-expression analysis, or proprietary affective outputs~\cite{nogueira2016vanishing,stein2018eeg,blom2019modeling,liang2023eeg,fisher2024exploring,shang2025general}. Similarly, studies in rehabilitation, exergaming, speech therapy, and audio-augmented reality often involved constrained populations, small samples, feasibility-oriented evaluations, or usability-focused assessments~\cite{hocine2015adaptation,pirovano2016intelligent,huber2021dynamic,loizou2025framework,giariskanis2025dynamic,shang2025general}. These studies remain valuable because they demonstrate complete-loop implementations in demanding applied contexts, but the evidence-maturity coding indicates that the strength of empirical support for the effects of adaptation varies across the corpus. The coding should therefore be read as a comparative evidence-maturity indicator, not as a standardized quality score.

Third, in some included studies, the distinction between player experience modeling and content adaptation was not clearly defined, complicating classification within the review framework. In Han et al.~\cite{han2024exploring}, gaming score relative to a real-time standard score both represents player ability and directly determines timer-change adaptation. Similarly, in Giariskanis et al.~\cite{giariskanis2025dynamic}, heuristic measures based on completion time and hint usage both characterize player skill and determine subsequent time and distance adaptations. Such methodological entanglements limited the consistent application of analytical categories across studies and highlight the need for clearer system descriptions in future work.

\subsection{Recommendations and Future Directions}
The findings point to three main directions for future work. First, future complete-loop systems should make the adaptation objective explicit, especially when affective information is used. Systems that use affective, physiological, or experiential information to support DDA, engagement, rehabilitation, or performance calibration should be distinguished from systems that treat affective state itself as the adaptation objective. This distinction is particularly important for stress, anxiety, horror, and related affective states, which were rarely treated as explicit adaptation targets within the selected corpus.

Second, future systems should explore input modalities that are less represented in the selected corpus but relevant to affective inference, including facial expression analysis and peripheral interaction data. These modalities may support less intrusive affect-informed or affective adaptation, provided that their interpretation is reported transparently and evaluated in relation to the intended adaptation objective~\cite{yang2021review,giannakakis2019review}.

Third, content-generation and content-modification methods, including PCG, should be reported as part of the adaptive loop rather than only as mechanisms for content variation. In complete-loop systems, these methods should be explicitly connected to the player experience model and to the adaptation objective. This would support clearer implementations of experience-driven content adaptation while preserving the distinction between player experience modeling and content adaptation~\cite{yannakakis2011experience}.

As an additional implementation direction, emerging generative methods, including Large Language Models (LLMs), may provide mechanisms for adaptive content generation, such as dialogue, narrative, or scenario adaptation. However, LLMs should be treated as content-generation mechanisms rather than as complete adaptive systems in themselves. Their use still requires clear player experience modeling, explicit adaptation objectives, empirical evaluation, and ethical safeguards~\cite{vartinen2024generating,yannakakis2023affective}. Furthermore, their use in complete-loop systems also raises practical challenges related to computational cost, response consistency, controllability, interpretability, and user safety.

\subsection{Ethical Considerations}
Ethical considerations apply to affect-adaptive systems regardless of the specific adaptation method. Systems that infer player experience or target affective states through adaptation should address transparency, informed consent, and the handling of affective or physiological data, particularly in health, rehabilitation, or other vulnerable-user contexts. These concerns become especially important when systems deliberately aim to induce, regulate, sustain, or attenuate states such as stress, anxiety, fear, or frustration.

Adaptation objectives should also be designed to avoid manipulative difficulty tuning or the reinforcement of negative affective states. When generative or affect-sensitive AI components are used, authors should report how consistency, controllability, interpretability, and user safety are addressed~\cite{yannakakis2023affective}.

\section{Conclusion}
\label{sec:conclusion}
This review analyzed 23 empirical studies published from January~1,~2015 to December~31,~2025 that implemented a complete experience-driven loop, defined as player data acquisition, player experience modeling, and adaptive game content. The review focused on how affective information is operationalized within these systems, with particular attention to whether stress, anxiety, horror, and related affective states are treated as explicit adaptation objectives.

Within the retrieved corpus, complete-loop implementations were relatively uncommon. Within the selected corpus, most systems were oriented toward DDA, engagement, rehabilitation, or performance-related objectives. Game telemetry was the dominant input modality, and knowledge-based approaches were the most common modeling and adaptation methods. Input modalities with affective relevance, such as facial expression analysis and peripheral interaction data, were less common within the selected corpus.

The main finding is that affective information entering an adaptive loop should not be equated with affective adaptation. Several systems used affective, physiological, or experiential cues, but these cues often supported challenge calibration, engagement maintenance, rehabilitation progress, or performance regulation. Far fewer systems treated affective state itself as the primary objective of adaptation. This distinction is especially important for stress, anxiety, horror, and related states, which were rarely addressed as explicit adaptation targets in complete empirically evaluated systems.

The screened and excluded records suggest that this scarcity reflects both fragmentation across sensing, modeling, and adaptation work and the practical difficulty of integrating these components into a cohesive adaptive loop. Addressing this gap would support the development of adaptive game systems that treat affective state as an explicit design objective, while also encouraging clearer reporting of player experience modeling, content adaptation, and empirical evidence for adaptive effects.

\section*{Acknowledgment}

The authors disclose that ChatGPT was used in the preparation of this manuscript. Its use was limited to restructuring existing text, performing spell checking, providing feedback on the organization and clarity of the paper, minimizing the redundant expansion of acronyms, and reducing unnecessary repetition of information. ChatGPT was not used to select studies, conduct the analysis, generate empirical findings, or create figures. All substantive content, analysis, and figures presented in this paper were produced independently by the authors.

\appendices
\section{Glossary}
\label{app:glossary}

\hangpara{2em}{1}\textbf{A/B Testing} An experimental method that compares two variants to determine which performs better. 

\hangpara{2em}{1}\textbf{Affective Modeling} The process of estimating a person's affective state from observable data. 

\hangpara{2em}{1}\textbf{Answer Set Programming} A declarative logic programming method for finding solutions that satisfy rules and constraints. 

\hangpara{2em}{1}\textbf{Bayesian Optimization} A sample-efficient optimization method that uses a probabilistic model to guide the search for good parameter settings. 

\hangpara{2em}{1}\textbf{Csikszentmihalyi's Flow Model} A theory of optimal experience in which engagement is supported by a balance between perceived challenge and skill~\cite{csikszentmihalyi1990flow}.

\hangpara{2em}{1}\textbf{Deep Reinforcement Learning} Reinforcement learning that uses neural networks to learn from complex data or decision spaces.

\hangpara{2em}{1}\textbf{Dynamic Difficulty Adjustment} The automatic modification of difficulty in response to a user's performance, state, or experience.

\hangpara{2em}{1}\textbf{Frijda's Model of Emotion} A theory that explains emotions as responses to events appraised in relation to a person's concerns, goals, or well-being~\cite{frijda1987emotion}.

\hangpara{2em}{1}\textbf{Fuzzy Methods} Methods that represent categories or rules with degrees of membership rather than strict yes-or-no boundaries. 

\hangpara{2em}{1}\textbf{Gradient Ascent Optimization} An optimization method that iteratively changes parameters in the direction that increases a chosen objective. 

\hangpara{2em}{1}\textbf{International Affective Picture System} A standardized image set used in emotion research, with normative ratings for valence, arousal, and dominance~\cite{lang1999iaps}.

\hangpara{2em}{1}\textbf{Markov Chains} Probabilistic models in which the next state depends on the current state through transition probabilities. 

\hangpara{2em}{1}\textbf{Matrix Factorization} A method that decomposes a matrix into lower-dimensional factors to reveal latent patterns. 

\hangpara{2em}{1}\textbf{Monte Carlo Tree Search} A tree-search method that evaluates possible decisions by simulating many future outcomes. 

\hangpara{2em}{1}\textbf{Multilayer Perceptron} A simple neural network made of several densely connected layers that learns patterns in data. 

\hangpara{2em}{1}\textbf{Partially Ordered Set Master} An online algorithm for DDA that maintains beliefs over ordered difficulty settings and updates them from feedback on whether a setting is too easy, too difficult, or appropriate~\cite{missura2011predicting}.

\hangpara{2em}{1}\textbf{Procedural Content Generation} The algorithmic creation of content. For example, it may generate levels, tasks, or game scenarios.

\hangpara{2em}{1}\textbf{Random Forest Classifier} A machine learning classifier that combines multiple decision trees to assign data to categories. 

\hangpara{2em}{1}\textbf{Reinforcement Learning} A machine learning approach in which an agent learns actions through rewards or penalties received from interaction.

\hangpara{2em}{1}\textbf{Russell's Circumplex Model of Affect} A model that represents affective states along valence and arousal dimensions~\cite{russell1980circumplex}.

\hangpara{2em}{1}\textbf{Support Vector Regression} A machine learning method for predicting continuous values. 

\hangpara{2em}{1}\textbf{Tensor Factorization} A method that decomposes multi-dimensional data into lower-dimensional factors to reveal latent patterns. 

\section{Detailed Summaries of Included Studies}
\label{app:detailed-summaries}

Hocine et al.~\cite{hocine2015adaptation} present a platformer game for upper-limb rehabilitation, which adapts in-game tasks (pointing movements) based on patient ability. Input data for adaptation included performance statistics and hand movement data, which were used to build a 2D ability-zone profile through a digital pheromone-based approach. Using Monte Carlo Tree Search (MCTS), the system dynamically adjusts difficulty by generating task sequences for different difficulty strategies, while online transitions between scenes adjust difficulty based on patient performance to improve training outcomes and sustain motivation. The generated level layout determines target/enemy positions and connectors; different visual themes are used to enhance aesthetics and engagement while preserving the same mechanics. The study involved seven post-stroke patients, demonstrating improved movement amplitudes with DDA compared to incremental and random difficulty strategies. The study has no connection to horror and only minimal, indirect ties to stress or anxiety, as the game is designed to reduce fatigue/frustration through DDA and motivational elements.

Zook \& Riedl~\cite{zook2015temporal} discuss a content adaptation loop within a turn-based role-playing game (RPG) in which players engage in spellcasting battles. The system collects player performance data, models behaviors using Tensor Factorization (TF), and optimizes game content with Answer Set Programming (ASP) to align gameplay with designer-defined performance curves. The study involved 32 players initially and 30 in a follow-up, showing TF's strong predictive power. The adaptation proved modestly effective in dynamically matching player abilities while highlighting challenges in addressing novel player learning behaviors. The findings suggest the potential of TF and ASP for proactive game content adaptation, with indirect relevance to stress through difficulty increases and possible frustration.

The study by Nogueira et al.~\cite{nogueira2016vanishing} integrates biofeedback mechanisms into \textit{Vanish}, a procedural horror game. It uses physiological data, including HR, EDA, and facial EMG, from 24 participants to estimate arousal and valence in real time. These estimates drive rule-based adaptations to level generation, creature encounters, audiovisual effects, and player attributes. Two biofeedback modes, symbiotic (explicit) and equilibria (implicit), are compared with a non-biofeedback control. Results show that biofeedback significantly affected key horror-relevant experience dimensions, including immersion, tension, positive affect, and negative affect.

In the study by Nagle et al.~\cite{nagle2016towards}, 160 participants played a custom first-person shooter (FPS) where players killed zombies. Personality traits, obtained through the Ten-Item Personality Inventory Questionnaire~\cite{gosling2003very}, were used to predict enjoyment and gameplay duration via linear regression. The regression informed the selection of the most suitable difficulty adaptation condition from both static difficulty curves and DDA. A secondary study (80 participants) showed that personality-based predictions significantly increased enjoyment and playtime compared to a default DDA approach. According to the authors, the results suggest that personality-driven adaptation can increase player engagement. Despite the game's theme, the study did not specifically address stress or anxiety in players.

The Intelligent Game Engine for Rehabilitation, proposed by Pirovano et al.~\cite{pirovano2016intelligent}, aims to improve rehabilitation exergames with adaptive and monitoring functionalities for safe and effective home therapy. Therapists configure initial parameters such as difficulty levels, motion ranges, and monitoring levels. The system then uses motion data (e.g., from Kinect or balance boards) and player performance during play to guide exercises such as weight shifting and lateral stepping. A fuzzy-based system monitors movement correctness, while Bayesian optimization adjusts game difficulty in real time. A virtual therapist avatar provides guidance and feedback. Tested with seven elderly participants, the system maintained an appropriate challenge level and was reported as engaging, with little reported stress.

Stein et al.~\cite{stein2018eeg} investigate EEG-triggered DDA in the multiplayer shooter game \textit{Boot Camp}. Excitement levels, detected by the Emotiv EPOC headset, are processed using thresholding, while game telemetry informs the game's difficulty adjustments, such as reducing damage for weaker players or increasing challenges for stronger players. Twenty-four participants played sessions with no DDA, heuristic DDA, and EEG-triggered DDA, with the latter showing improvements in long-term player excitement and engagement over the alternatives. While the study focuses on enhancing player enjoyment and engagement, it has only a minor indirect relation to stress or anxiety through the inherent challenges of competitive gaming.

Targeting flow state maintenance, Bicho \& Martinho~\cite{bicho2018multi} developed a side-scrolling platformer game with multi-dimensional level challenges dynamically generated according to player skill progression in various game mechanics. The player experience model is updated using a weighted rolling window to track recent performance, while player preferences were evaluated through a form of A/B testing that presented players with challenges adapted to their strongest and weakest skills. The study involved 30 participants and demonstrated consistent player preference for adapted challenges. This work emphasized balancing difficulty to reduce frustration and maintain engagement, with minimal relation to horror and only indirect connections to stress or anxiety through challenge calibration.

The work by Mostefai et al.~\cite{mostefai2019generic} proposes an emotion-driven approach for an educational programming game. The game's content is adapted in real time to sustain engagement by promoting positive affective states and reducing negative or disengaging states such as frustration and boredom. A preliminary player profile, including personality type and playing style, is initially obtained via questionnaires. During gameplay, players perform program evaluation tasks, and their actions and goal statuses, together with this profile, are used to determine their current emotional state using a rule-based approach grounded in Frijda's Model of Emotion~\cite{frijda1987emotion}. This state is then used with Markov transition matrices relating game events to emotional changes, enabling the adaptation of program tasks, difficulty, and pacing through heuristic rules. Thirty university students (15 experimental, 15 control) participated, with the experimental group showing higher engagement, more positive affective responses, fewer negative or disengaging states, and better learning outcomes. This study addresses stress indirectly but does not focus on horror or anxiety.

Blom et al.~\cite{blom2019modeling} analyzed DDA using facial behavior (FB), including facial expressions and head pose. The authors evaluated adaptive level-chunk generation in a \textit{Super Mario} clone using frame-wise emotion estimates from a proprietary FB software toolkit. The first adaptive experiment (10 participants) used a heuristic approach that averaged emotion probability estimates over time and applied Gradient Ascent Optimization to adjust chunk difficulty according to predefined emotion-based rules. A subsequent training phase (25 participants, no online content adaptation) trained a Random Forest Classifier (RFC) to estimate chunk-level player challenge from detected emotions, head pose, current difficulty, and user feedback labels. The final experiment, also with 25 participants, used the trained RFC to adjust game content in real time. Results showed that players generally preferred the personalized system, especially from easy and normal starting difficulties. The model-based approach converged faster than the heuristic one, while frustration effects were mixed, with one harder condition rated as especially frustrating. The study addressed stress-related frustration through DDA but did not explore horror or anxiety.

Sifa et al.~\cite{sifa2020matrix} propose adapting quest recommendations in an online RPG using matrix factorization (MF) and TF models to profile player skills and preferences based on players' in-game behavioral metrics, such as health lost and completion time. These models recommend personalized quests from a pool of procedurally generated options, selecting those that match the player's profile to balance challenge and engagement. Evaluation with 25$\,$686 players showed that TF achieved the best retention and abandonment outcomes, outperforming random, rule-based, and neighborhood-based approaches, while MF strongly reduced quest failure rates. While not targeting horror, stress, or anxiety, the study has a weak indirect relation through challenge balancing.

The paper by Pfau et al.~\cite{pfau2020enemy} examines a content adaptation loop in solo runs of massively multiplayer online RPG dungeons featuring both heuristic-based and neural network-based enemies. For heuristic-based enemies, input data includes performance metrics (e.g., successful completions, deaths, timeouts), leading to fixed difficulty adjustments such as spawn rates and health points. For neural network-based enemies, input data consists of detailed player behavioral logs (e.g., skill rotations, situational responses) used to train a multi-layer perceptron, that is, a standard neural network that emulates player strategies. In a four-week study with 171 participants, neural network-based enemies supported stronger long-term engagement; among the 30 questionnaire respondents, they significantly outperformed heuristic ones in fostering intrinsic motivation, particularly in terms of interest-enjoyment, while slightly increasing stress-related tension-pressure compared to some heuristic enemies.

Lara-Alvarez et al.~\cite{lara2021induction} developed and tested a fuzzy logic-based educational math game that adapts arithmetic difficulty and sound effects based on speech-derived emotional states and performance data to improve learning by reducing unpleasant, anxiety-related states. Emotional states (valence and arousal) are modeled using Russell's Circumplex Model of Affect~\cite{russell1980circumplex} and a pre-trained support vector regression model trained on a labeled speech-emotion dataset~\cite{perez2011emowisconsin}, while fuzzy control adjusts difficulty and aesthetic content to promote positive emotions. In a study with 40 students, the system achieved faster transitions from negative low-arousal emotional states toward more positive or activated states compared to a linear DDA baseline. The method addresses anxiety-related affective regulation in educational gameplay but does not address horror.

The study by Huber et al.~\cite{huber2021dynamic} investigates dynamic content adaptation in a VR exergame where players navigate mazes containing predefined physical exercise tasks. At the end of each maze, players provide subjective difficulty ratings, which are used as input for a Deep Reinforcement Learning (DRL)-based system to adapt maze complexity and exercise room selection dynamically. The DRL system is pre-trained via simulation before being fine-tuned during gameplay. The study involved 19 participants, with results indicating promising difficulty adaptation, particularly for players who found the initial maze too easy or too hard. The research emphasizes balancing cognitive and physical challenges to improve player engagement, with no direct connection to feelings of horror, stress, or anxiety.

Rosa et al.~\cite{rosa2021dynamic} tested a hybrid DDA system in a 2D platformer game using performance metrics (e.g., deaths and jump success/failure) and affective data derived from EDA. The EDA signal was processed into tonic and phasic components to estimate arousal, which, together with performance data, informed heuristic DDA models. The system adapted platform sizes and distances, as well as player jump height, to tailor difficulty and support flow. Across three experiments with 20, 36, and up to 132 participants, the DDA variants increased level completion and reduced performance dispersion, with the hybrid model showing better perceived difficulty adequacy and especially positive effects for players preferring easier games. The study primarily focused on DDA, engagement, and flow, with only minimal indirect relation to horror, stress, or anxiety.

Chen et al.~\cite{chen2023impact} present a complete content adaptation loop in a walking simulator with the goal of improving immersion by measuring hemodynamic responses in players via functional near-infrared spectroscopy (fNIRS). User and data modeling was performed using rule-based calculations of focus levels from oxygenated and deoxygenated hemoglobin, which informed the adaptation of related visual effects such as depth of field, color grading, and chromatic aberration. Among 24 participants, the Brain Computer Interface (BCI)-adaptive version slightly increased immersion compared with a non-BCI version in which visual effects were triggered by the player's in-game actions, such as movement patterns, with significant improvements in temporal dissociation. The focus of this study was on creating an immersive experience using realistic effects that mimic focus loss, without a direct relationship to horror, stress, or anxiety.

In turn, Liang et al.~\cite{liang2023eeg} discuss an EEG-driven system for emotional regulation, where users explore a VR scenario. The EEG device outputs real-time attention and relaxation levels, and a rule-based system adapts affective scene elements, such as butterflies, puppies, kittens, wolves, and bears, selected with reference to the International Affective Picture System (IAPS)~\cite{lang1999iaps} and its valence, arousal, and dominance ratings. Eighty college students participated in a comparison between relaxation imagery, 2D pictures/videos, non-adaptive VR, and EEG-adaptive VR. Self-reported emotional ratings and post-experiment user experience questionnaires suggest that the adaptive VR condition produced more stable emotion regulation and higher user satisfaction than the alternatives. The authors frame the system as a promising approach for personalized stress reduction and emotional relaxation, although further evaluation is needed.

Fisher and Kulshreshth~\cite{fisher2024exploring} present an FPS game in which players survive alien attacks while escorting cows, collecting crystals, and destroying UFOs. In a study with 31 participants, including 14 casual/non-gamers and 17 experienced gamers, the system collected gameplay metrics plus EEG stress/engagement and HR data, then models player state through weighted and moving averages, normalized affect aggregation, and rule-based difficulty mapping. The game adapts difficulty in real time by modifying player and enemy parameters, including aim assist, reload speed, enemy health, fire rate, and spawn behavior. Results showed no single DDA method clearly outperformed static difficulty overall, though emotion-based DDA aligned with Flow theory~\cite{csikszentmihalyi1990flow} and directly addressed stress, with only indirect relation to anxiety and none to horror.

Han et al.~\cite{han2024exploring} present an immersive VR soccer exergame in which players act as goalkeepers and block incoming balls with virtual hands. The DDA mechanism uses gameplay performance, specifically gaming score, and compares it with a real-time standard gaming score to estimate player ability through a rule-based procedure. Emotion is measured only as a secondary outcome and is not used as input to the adaptation mechanism. The score comparison drives timer-change adaptation every 10 seconds, altering time pressure and thereby the effective task difficulty. The study involved 14 university participants with a mean age of 25.4 years, although the reported gender breakdown appears internally inconsistent. Participants played both adaptive and non-adaptive versions. The timer-change condition was associated with significantly more neutral emotions and higher heart rate, as well as a slightly higher but non-significant gaming score, suggesting improved challenge calibration. The paper has no horror focus and only indirect links to stress or anxiety through challenge calibration, comfort, and frustration avoidance.

Loizou \& Andreou~\cite{loizou2025framework} propose a domain-agnostic framework for real-time self-adaptive serious games using semantic annotation, expert-defined rules, and a digital twin/rules engine mechanism. The framework is demonstrated with two games: a speech therapy game for children with learning difficulties or syndromes, focused on phonological awareness and syllable processing tasks, and an industrial training game for poultry factory engineers and staff, focused on machinery diagnosis and climate control. However, the only clearly reported human-player demonstration of the complete adaptation loop is the speech therapy game, where five children up to 12 years old completed a one-hour supervised syllable recognition task. In this game, gameplay telemetry such as accuracy, response time, errors, and progression is semantically annotated, compared against expert-defined thresholds, and used to adapt syllable count, word complexity, hints, visual and sound-based feedback cues, and timing. A broader evaluation with 15 domain experts from therapeutic and industrial contexts assessed the framework positively, mainly in terms of perceived usability, adaptability, and expert satisfaction. The study has no horror focus, and its relation to stress or anxiety is only indirect, through challenge calibration and the framework's proposed capacity to incorporate biosignals.

Giariskanis et al.~\cite{giariskanis2025dynamic} present a head-worn audio augmented reality serious game set in a museum, where players locate, collect, and transfer virtual artifacts using spatial audio cues. Player performance telemetry, especially completion time, remaining or additional time, and hint usage, drives a heuristic DDA mechanism that adapts next-round difficulty by changing available time and artifact visibility distance. The adapted scenario therefore remains largely the same, but its challenge is personalized between rounds. The study involved 20 participants aged 16--54, evaluating both interaction usability and DDA behavior. Results suggest that the mixed interaction mode was more usable than voice-only and gaze-only interaction, although not significantly different from hand-only interaction, and observational DDA results showed that allocated task times closely matched players' actual completion times after adaptation. The paper supports engagement and flow-oriented difficulty balancing, but has no direct horror focus and only indirect links to stress or anxiety.

Croissant et al.~\cite{croissant2025advancing} present a complete affect-adaptive loop in a bespoke 2D arcade-style action game where players avoid projectile attacks using movement, blocking, and dashing. In Experiment 1, 161 participants provided self-reported valence/arousal and gameplay telemetry, enabling correlation and linear regression modeling of player state and tests of manipulated game variables, especially enemy attack speed; musical tempo and enemy synchronization were also tested as candidate output variables, but these were not carried forward into Experiment 2. In Experiment 2, 158 participants were assigned to control, level-based adaptation, or player-based adaptation. The game adapted enemy attack speed using either level context or a regression-based player experience model derived from blocks and deaths. Player-based adaptation significantly increased self-reported valence relative to both control and level-based adaptation. The study is primarily affect-focused, with only indirect relevance to stress/anxiety and no specific relation to horror.

The study by Mario \& Arifin~\cite{mario2025analyzing} presents an in-progress, turn-based survival roguelike chess prototype in which players control a knight and survive increasingly difficult rounds of enemy pieces. The system uses gameplay telemetry, including damage dealt and received, current health, powerup usage, decision time, and movement efficiency, to estimate player performance through normalized indices combined via weighted linear combination. This score drives DDA by procedurally parameterizing enemy attributes, visual hints, powerup price and quality, and move time limits. Preliminary testing involved 14 beta testers playing local copies of the prototype on their own devices and completing the User Experience Questionnaire (UEQ)~\cite{laugwitz2008construction} and custom Likert-scale questionnaires. Results were broadly positive, with favorable UEQ scores and perceived challenge impact, although the evidence remains preliminary and self-report based. The study has no direct horror focus and only indirect links to stress or anxiety through difficulty balancing and flow-oriented challenge regulation.

Finally, Shang et al.~\cite{shang2025general} present a haptic/biometric DDA loop for robot-assisted upper-limb rehabilitation games. Patients interact with therapy tasks through a haptic robot, generating gameplay, robotic/kinematic, HR, and EDA data. A modified Partially Ordered Set Master algorithm with belief vectors~\cite{missura2011predicting} combines performance observations with physiological stress estimates to infer an appropriate challenge level. The system adapts game parameters, haptic force feedback, guidance, and audiovisual/text feedback in three upper-limb rehabilitation exergames targeting different robot-mediated motor control tasks. Eleven post-stroke participants with mild-to-moderate motor and cognitive impairments tested games with and without DDA. Participants favored DDA trials, with 36\% preferring DDA versus 7\% non-DDA on average for the three games, while 57\% reported no preference. The study relates to stress through biometric modeling, but not horror.

\ifCLASSOPTIONcaptionsoff
  \newpage
\fi

\bibliographystyle{IEEEtran}

\begin{IEEEbiography}[{\includegraphics[width=1in,height=1.25in,clip,keepaspectratio]{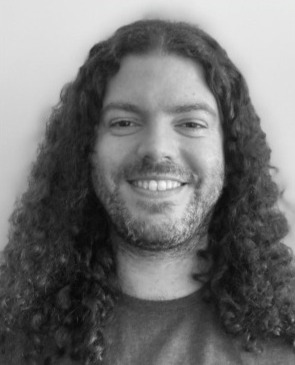}}]{Phil Lopes} holds a Ph.D. degree in Artificial Intelligence in Games from the Institute of Digital Games of the University of Malta (2017), and a B.Sc. and M.Sc. in Computer Science from the Faculty of Sciences of the University of Lisbon, Portugal (2009 and 2012, respectively). For four years following his Ph.D., Dr. Lopes was a Post-Doctoral Researcher at the University of Geneva and the \'{E}cole Polytechnique F\'{e}d\'{e}rale de Lausanne (EPFL), in Switzerland. He is currently an Assistant Professor at Lusófona University in Lisbon, where he also serves as the Director of the Master’s program in Artificial Intelligence for Games. Dr. Lopes is an Associate Editor for the IEEE Transactions on Games journal and a member of the Editorial Board of the journal Cyberpsychology, Behavior, and Social Networking. His research interests include Procedural Content Generation, Affective Computing, and Emotion Recognition, with applications in digital games, virtual environments, and human-computer interaction.
\end{IEEEbiography}

\begin{IEEEbiography}[{\includegraphics[width=1in,height=1.25in,clip,keepaspectratio]{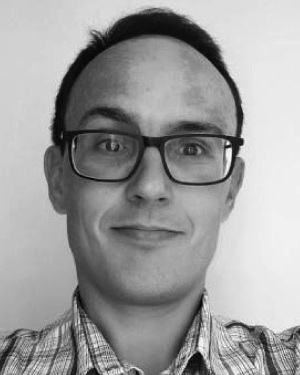}}]{Nuno Fachada} received the B.Sc., M.Sc., and
Ph.D. degrees in Electrical and Computer Engineering from IST/Universidade de Lisboa, Portugal, in 2005, 2008, and 2016, respectively. After a year
as a Postdoctoral Researcher at IST, he moved to Lusófona University, where he is an Assistant Professor at the School of Communication, Arts and Information Technology, where he teaches in the Videogames BA and in the AI for Games MSc programs, and a Researcher at INESC INOV-Lab. Nuno is currently an Associate Editor at the Journal of Open Research Software, and his main research interests include artificial intelligence, computer games, modeling and simulation, research software, and computer science education.
\end{IEEEbiography}

\begin{IEEEbiography} [{\includegraphics[width=1in,height=1.25in,clip,keepaspectratio]{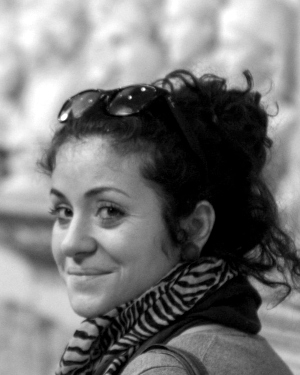}}]{Maria Micaela Fonseca}
holds a BSc in Physics Engineering and a Ph.D. in Physics (2011), both from NOVA University Lisbon, where she received the Best Student Award in Physics Engineering. She is Vice-Director of the HEI-Lab research unit – Digital Human-Environment Interaction Lab at Lusófona University, where she leads interdisciplinary initiatives bridging science, technology, and creative practice. Her research focuses on serious games and affective computing. Micaela is the group leader of the Wellbeing Interactivity Nexus Lab and co-coordinates the VR development team at HEI-Lab. She co-founded both the Games and Social Impact Media Research Lab and the International Journal of Games and Social Impact, where she currently serves as Editor.
\end{IEEEbiography}

\end{document}